# Detection of antiskyrmions by topological Hall effect in Heusler compounds

Vivek Kumar,[1] Nitesh Kumar,[1] Manfred Reehuis,[2] Jacob Gayles,[1] A. S. Sukhanov,[1,3] Andreas Hoser,[2] Françoise Damay,[4] Chandra Shekhar,[1] Peter Adler,[1] and Claudia Felser[1*]

[1]*Max Planck Institute for Chemical Physics of Solids, 01187 Dresden, Germany*
[2]*Helmholtz-Zentrum Berlin für Materialien und Energie, 14109 Berlin, Germany*
[3]*Institut für Festkörper- und Materialphysik, Technische Universität Dresden, 01069 Dresden, Germany*
[4]*Laboratoire Léon Brillouin, CEA-CNRS, CEA Saclay, 91191 Gif-sur-Yvette, France*

**Heusler compounds having $D_{2d}$ crystal symmetry gained much attention recently due to the stabilization of a vortex-like spin texture called antiskyrmions in thin lamellae of $Mn_{1.4}Pt_{0.9}Pd_{0.1}Sn$ as reported in the work of Nayak *et al.* [Nature 548, 561 (2017)]. Here we show that bulk $Mn_{1.4}Pt_{0.9}Pd_{0.1}Sn$ undergoes a spin-reorientation transition from a collinear ferromagnetic to a non-collinear configuration of Mn moments below 135 K, which is accompanied by the emergence of a topological Hall effect. We tune the topological Hall effect in Pd and Rh substituted $Mn_{1.4}PtSn$ Heusler compounds by changing the intrinsic magnetic properties and spin textures. A unique feature of the present system is the observation of a zero-field topological Hall resistivity with a sign change which indicates the robust formation of antiskyrmions.**

## I. INTRODUCTION

Topological magnetic textures show promise in many areas of technology, such as racetrack memory [1,2] and neuromorphic computing [3,4], due to the unique transport properties that allow for efficient manipulation [5,6] and detection [7,8] of the magnetic textures. This is a consequence of the real space Berry curvature of the topological texture that can be tuned by the electronic structure [9,10]. One such topological texture is the skyrmion [11,12] which is found in chiral magnets [13] with a unit integer topological charge. Recently antiskyrmions (aSKs) [14], the anti-particles of the skyrmions, were stabilized in thin lamellae of the tetragonal inverse Heusler compound $Mn_{1.4}Pt_{0.9}Pd_{0.1}Sn$ for a broader field range and even above room temperature [15]. In this compound ($D_{2d}$ symmetry class), the Dzyaloshinskii-Moriya interaction (DMI) is anisotropic where the DM vectors along the *x* and *y* directions in the basal plane are of opposite sign and hence cause the stability of the aSKs [16]. The in-plane winding of the spin texture for an aSK is anisotropic with an opposite topological charge as compared to that of the azimuthally symmetric Bloch and Néel skyrmion of the same charge [17]. The topological winding induces a real space Berry curvature that strongly influences all transport properties [18]. The topological Hall effect (THE) is one such electrical transport property that has been intrinsically linked to the skyrmions [8].

The THE results in a further component of the total Hall signal, in addition to the normal Hall and the magnetization-scaled anomalous Hall components [19]. It arises due to the emergent magnetic field which is a result of the finite Berry phase contributions from either real or momentum spaces when the conduction electrons interact with a non-coplanar (NCP) spin texture of finite chirality [20]. The THE was widely studied in the skyrmions hosting B20 compounds [7,8,19,21–24]. It was also observed in other systems possessing NCP spin structures *e.g.* pyrochlore lattices [25,26], antiferromagnets [27,28], spin glasses [29,30] and correlated systems such as oxide thin films [31,32].

A chemical substitution in place of the heavy element Pt in the non-centrosymmetric Heusler compound $Mn_{1.4}PtSn$ can allow for systematic control of the intrinsic properties that give rise to topological textures, namely spin orbit coupling (SOC), electron occupation and magnetization. These properties determine fundamental parameters such as the exchange interactions, the DMI, and the magnetocrystalline anisotropy (MCA). In turn, the competition of these parameters determines which magnetic texture is formed, *i.e.* a long wavelength spiral or a topological spin texture. The presence of more than one magnetic sublattice and the competition between parallel and antiparallel exchange interactions in the non-centrosymmetric crystal can lead to a spin reorientation transition [33]. In $Mn_{1.4}PtSn$ below 160 K the formation of a non-collinear spin structure was reported [15]. In this article, we investigate the detailed magnetic structures of $Mn_{1.4}Pt_{0.9}Pd_{0.1}Sn$ below and above the spin reorientation transition temperature $T_{SR}$ and demonstrate that below $T_{SR}$ a THE emerges with features which are evidence for the presence of aSKs in the bulk material. Fractional substitution of Pt by Pd (isoelectronic substitution) in $Mn_{1.4}PtSn$ is used to exclusively change the SOC, whereas substitution by Rh varies the electron occupancy of the bands and in consequence changes the SOC as well as the magnetization. Our results demonstrate the tunability of topological spin textures in the Heusler system which paves the way for systematic design of aSK phase containing compounds suitable for applications in spintronics. The THE can be used as a sensitive probe for characterizing the spin textures.

## II. EXPERIMENT

Polycrystalline ingots of $Mn_{1.4}Pt_{1-x}Pd_xSn$ ($x$ = 0, 0.1 … 0.3) and $Mn_{1.4}Pt_{1-y}Rh_ySn$ ($y$ = 0.1, 0.2 … 0.8) with the stoichiometric amounts of constituent elements were prepared by arc-melting in the presence of Ar atmosphere. The synthesized ingots were sealed in an evacuated quartz tube and annealed at 873 K (Pd-substituted samples) and 1073 K (parent compound $Mn_{1.4}PtSn$ and Rh-substituted samples) for 7 days followed by quenching into an ice-water mixture. The phase purity and the crystal structure of the samples were examined by room temperature powder x-ray diffraction (XRD) using a Huber G670 camera [Guinier technique, $\lambda$ = 1.54056 Å (Cu−K$\alpha_1$ radiation)].

Neutron powder diffraction experiments of $Mn_{1.4}Pt_{0.9}Pd_{0.1}Sn$ were carried out on the instrument E6 at the BER II reactor of the Helmholtz-Zentrum Berlin, where powder diffraction patterns were recorded between the diffraction angles ($2\theta$) 5.5 and 136.5°. This instrument uses a pyrolytic graphite (PG) monochromator selecting the neutron wavelength $\lambda = 2.43$ Å. For the experiment the powder sample was filled in a vanadium container (cylindrical) of diameter 6 mm and height 4 cm. The temperature dependence of the crystal and magnetic structure was investigated between 2.4 and 449 K using an Orange Cryofurnace (AS Scientific Products Ltd., Abingdon, GB). In order to follow in detail the structural and magnetic changes, we collected in this temperature range 47 powder diffraction patterns. Data sets at 2.4, 203 and 449 K, were measured with an increased counting rate in order to determine the positional parameters and magnetic moments with good accuracy.

Neutron powder diffraction measurements of $Mn_{1.4}Pt_{1-y}Rh_ySn$ ($y = 0.4, 0.6, 0.7$ and $0.8$) were taken on the cold neutron two-axis powder diffractometer (instrument G 4-1) at Laboratoire Léon Brillouin, France. The neutron wavelength $\lambda = 2.43$ Å was selected using a pyrolytic graphite (PG) monochromator. Powdered samples were filled in a vanadium container (cylindrical) of diameter 6 mm and height 4 cm. Samples were cooled down in a cryostat to 1.5 K and measurements were taken while warming in the $2\theta$ range between 2 and 82°. Rietveld refinements of all the powder diffraction data were carried out with the program *FullProf* [34], using the nuclear scattering lengths $b(Mn) = -3.73$ fm, $b(Pd) = 5.91$ fm, $b(Pt) = 9.63$ fm, and $b(Sn) = 6.228$ fm [35]. The magnetic form factor of the Mn atoms was taken from Ref. [36].

Magnetization measurements were performed using a vibrating sample magnetometer (MPMS3, Quantum Design, $\mu_0 H_{max} = 7$ T). The electrical transport properties were investigated using a physical property measurement system (PPMS9, Quantum Design, $\mu_0 H_{max} = 9$ T). The samples were cut into rectangular bar shaped pieces with dimensions of approximately $3.5 \times 1.5 \times 0.5$ mm$^3$. Point contacts were made on the samples with 25 μm diameter Pt wires and silver paint. The five probe geometry was used for the Hall experiments. The Hall resistivity was measured at different temperatures between 2 and 300 K in the field range from −5 to 5 T in hysteresis mode by applying the field perpendicular to the rectangular surface of the bar. In order to avoid any artifact due to the demagnetization effect, both magnetization and resistivity measurements were performed on the same piece (length × breadth × height) where the magnetic field was applied perpendicular to the rectangular surface (length × breadth) under identical conditions. It should be noted that in a superconducting coil trapped magnetic flux may lead to deviations of reported and real magnetic field. Especially, when measuring the magnetization, magnetoresistance, or the Hall resistivity of a soft ferromagnet in a field sweep through zero, the reversal of the hysteresis (i.e. of the coercive fields) may be observed [37]. The presented data were measured in a MPMS3 magnetometer with a 7 T magnet and a PPMS with a 9 T magnet, respectively. In field sweeps both systems typically produce a maximum field error of 20 Oe around zero field [37].

## III. RESULTS

### A. Crystal structure and lattice parameters

We determined the crystal structure and lattice parameters of the prepared samples by Rietveld refinement of the powder x-ray diffraction patterns shown in Fig. S1 of the Supplemental Material [38] using the *Fullprof* program. The tetragonal crystal structure is retained in both the substitution series $Mn_{1.4}Pt_{1-x}Pd_xSn$ and $Mn_{1.4}Pt_{1-y}Rh_ySn$. All samples crystallize in the inverse tetragonal Heusler structure (space group $I\bar{4}2d$, No. 122). The lattice constants, *c/a* ratios and the cell volumes of $Mn_{1.4}Pt_{1-x}Pd_xSn$ and $Mn_{1.4}Pt_{1-y}Rh_ySn$ are shown in Fig. 1. The lattice constant *a* increases continuously by Pd substitution whereas *c* gets reduced slightly compared to the parent compound and remains almost constant with increasing *x*, thereby the *c/a* ratio decreases and the cell volume increases. On the other hand, Rh substitution first increases *a* and decreases *c* up to $y = 0.5$ and thereafter a reverse trend is observed *i.e.*, *a* decreases and *c* increases from $y = 0.6$ to 0.8. Consequently, the *c/a* ratio decreases up to $y = 0.5$ and then increases considerably. However, the cell volume decreases linearly up to $y = 0.5$ and thereafter the decrement is non-monotonic.

In this system, the tetragonal structure results from two cubic unit cells slightly displaced along the *c* axis which leads to a *c/a* ratio close to 2. Hence, the degree of tetragonality increases when the *c/a* ratio tends away from the value 2 or in other words the *c/a* ratio moves away from the pseudo-cubic condition ($c = a'$, where $a' = 2a$). The decrease in the *c/a* ratio for all Pd substitutions and moderate Rh substitutions up to $y = 0.5$ suggests an enhanced tetragonality. Accordingly, higher Rh substitution, $y > 0.5$, leads to a decrease in tetragonality as reflected in the drastic rise in the *c/a* ratio.

### B. Neutron powder diffraction

We investigated the detailed magnetic structure of the present system using neutron-powder-diffraction experiments. In Fig. 2 the neutron-powder-diffraction patterns of $Mn_{1.4}Pt_{0.9}Pd_{0.1}Sn$ are shown, the corresponding Rietveld refinement results are summarized in Table S1 of Supplemental Material [38]. The data collected at 449 K (well above the Curie temperature, $T_C = 390$ K) was successfully refined in the space group $I\bar{4}2d$ (No. 122) as shown in Fig. 2(a). In this space group two different Mn atoms (labeled as Mn1 and Mn2) are located at the Wyckoff positions $8d(x,¼,⅛)$ with $x \sim 0.73$ and $4a(0,0,0)$ [in our case we used the second site generated by $(0,½,¼) = (½,½,½) + (½,0,¾)$], while the Sn atoms are located at $8d(x,¼,⅛)$ with $x \sim 0.28$. The refinements showed that the Mn2 atoms at the site $4a$ only reach an occupancy of 0.772(19). For the Pt and Pd atoms, which are statistically distributed at the position $8c(0,0,z)$ [in our case we used the second site generated by $(0,½,z) = (½,½,½) + (½,0,z'+¾)$] (with $z \sim 0.02$), one obtains the occupancies $occ(Pt) = 0.902(22)$ and $occ(Pd) = 0.098(22)$. This finally gives the formula $Mn_{1.39(1)}Pt_{0.90(2)}Pd_{0.10(2)}Sn$. Alternatively, the crystal structure of $Mn_{1.4}Pt_{0.9}Pd_{0.1}Sn$ has been described earlier within space group $I\bar{4}2m$ (No. 121) [15]. In this space group additional reflections (002, 110, 114, 222, 006, etc.) are generated which are not observed in the experimental patterns. Accordingly, $Mn_{1.4}Pt_{0.9}Pd_{0.1}Sn$ crystallizes in the space group $I\bar{4}2d$.

A previous report [15] and our magnetization measurements (see the Supplemental Material [38]) indicated the presence of a transition into a ferro- or ferrimagnetic state at the Curie temperature $T_C \sim 390$ K followed by a spin reorientation transition, which sets in at $T_{SR} < 150$ K. Therefore, in order to determine the high-temperature magnetic structure of $Mn_{1.4}Pt_{0.9}Pd_{0.1}Sn$, we used the data set collected at 203 K well between the two transition temperatures $T_C$ and $T_{SR}$. The strongest magnetic intensities are superimposed on the positions of the nuclear reflections 112, 200, 220 and 204 (shown in Fig. 2(b)), indicating a ferromagnetic ordering of the Mn atoms with a propagation vector $\mathbf{k} = 0$. Due to the fact that magnetic intensity is observable on the reflections 200 and 220 it can be assumed that the magnetic moments are predominantly aligned parallel to the $c$ axis. Therefore, we used a magnetic structure model where only the $z$ components of the magnetic moments $\mu_z(Mn1)$ and $\mu_z(Mn2)$ were allowed to vary, see Table S1 in Supplemental Material [38]. The magnetic moments were found to be $\mu_z(Mn1) = 3.32(3)\ \mu_B$ and $\mu_z(Mn2) = 1.70(6)\ \mu_B$. The excellent agreement between the observed and calculated intensity of the reflection 004 suggests that no additional ferromagnetic component occurs in the $ab$ plane at 203 K. The neutron powder diffraction analysis reveals that the high-temperature magnetic configuration has a collinear ferromagnetic spin alignment along the tetragonal axis, in contrast to the previous suggestion of a ferrimagnetic spin structure [15]. The proposed magnetic structure was based on the general model of $Mn_2$-based inverse Heusler compounds, which frequently have a ferrimagnetic configuration due to antiparallel spin alignment between Mn atoms at different sites [39,40]. A simulation of a ferrimagnetic model demonstrates that strong magnetic Bragg intensity is generated at the position of the relatively weak nuclear reflection 101 which is not observed.

Below 135 K, our neutron powder diffraction data showed a spontaneous increase of magnetic intensity on the reflections 101 and 004. The strong increase of the magnetic intensity of the 004 reflection indicates the presence of an additional spin alignment in the *ab* plane. We focused on the data set collected at 2.4 K (shown in Fig. 2(c)) to solve the ground state magnetic configuration. The refinements showed that it is impossible to determine the moment direction within the *ab* plane. Therefore, we have set the moment direction of Mn1 and Mn2 parallel to the *a* axis. The moments of the two Mn atoms are coupled antiparallel resulting in a ferrimagnetic ordering in the *ab* plane. At 2.4 K the magnetic moments reach in the plane the moment values $\mu_x$(Mn1) = 2.89(7) $\mu_B$ and $\mu_x$(Mn2) = −1.65(11) $\mu_B$. The observed magnetic structure is compatible with the symmetry of the space group $I\bar{4}2d$ as shown in Supplemental Material [38].

It is interesting to see that from 203 to 2.4 K the *z* component of Mn1 is strongly decreasing [from 3.32(3) $\mu_B$ to 2.56(6) $\mu_B$], whereas that of Mn2 is strongly increasing [from 1.70(6) $\mu_B$ to 3.13(7) $\mu_B$]. This shows that the moment of Mn1 is much stronger tilted from the *c* axis (48.5°) than that of Mn2 (27.7°). The total moments of the two manganese atoms are $\mu_{tot}$(Mn1) = 3.87(5) $\mu_B$ and $\mu_{tot}$(Mn2) = 3.54(5) $\mu_B$.

From Fig. 2(d), where the temperature dependence of the magnetic moments is shown, it is apparent that the spin reorientation transition at $T_{SR}$ leads to a spontaneous increase of the total moments of the atoms Mn1 and Mn2. The total magnetic moments derived from the neutron data compare well with those from the magnetization measurements, see Supplemental Material [38]. From the temperature dependence of the moments the transition temperatures $T_C$ = 390(4) K and $T_{SR}$ = 135(3) K are obtained. In Fig. 2(e) and (f) we show the thermal variation of the lattice parameters *a* and *c*, and the cell volume, *V*, which are continuously decreasing to 2.4 K. Both magnetic transitions are reflected in slight anomalies, which is most apparent for the *c*/*a* ratio (shown in Fig. 2f).

Our neutron investigations demonstrate that Mn$_{1.4}$Pt$_{0.9}$Pd$_{0.1}$Sn displays a collinear ferromagnetic spin alignment along the tetragonal axis in the temperature range $T_{SR} < T < T_C$, whereas the magnetic structure transforms into a non-collinear spin configuration for $T < T_{SR}$. In Fig. 3 the magnetic structures below (2.4 K) and above (200 K) $T_{SR}$, as well as the angle between the Mn1 and Mn2 moments are shown. The spin reorientation transition involves spin canting of each Mn sublattice as well as an increase in the total moments. These findings are consistent with the temperature dependent magnetization data presented in Fig. S3 of Supplemental Material [38]. The low-temperature spin structure is non-collinear, but still coplanar. Recently, a non-coplanar spin arrangement has been discussed as a candidate structure for Mn$_{1.4}$PtSn even at zero-field, where an additional antiferromagnetic component may be present parallel to *y* [41]. This order should generate magnetic intensity at the position of the reflection 002. From our data of Mn$_{1.4}$Pt$_{0.9}$Pd$_{0.1}$Sn collected at 2.4 K we were not able to find significant magnetic intensity at this position. Simulations showed that any magnetic moment parallel to *y* is expected to be smaller than 0.2 $\mu_B$.

Neutron patterns of Mn$_{1.4}$Pt$_{0.6}$Rh$_{0.4}$Sn verify that for moderate degrees of Rh substitution a non-collinear spin structure and collinear ferromagnetic structure are retained below and above $T_{SR}$, respectively, whereas for higher substitution levels $y > 0.5$ an essentially collinear ferrimagnetic (FiM) spin configuration with both Mn moments lying in the basal plane emerges at all temperatures below $T_C$ (see the supplemental Material [38]). The change in spin structure is associated with a decrease in the tetragonal distortion of the crystal structure for $y > 0.5$ as discussed in Section III.A. (*c.f.* Fig. 1). The neutron diffraction data, however, cannot exclude the persistence of a small spin canting which might lead to a topological Hall effect in an external magnetic field.

### C. Magnetic properties

The evolution of the magnetic properties for both substitution series is depicted in Fig. 4 and Figs. S4 and S5 of Supplemental Material [38]. The parent sample exhibits a $T_C$ of ~ 400 K which reduces to 375 K by increasing the Pd substitution to $x = 0.3$. On the other hand, Rh substitution results in a $T_C$ of 246 K for $y = 0.8$. The magnetization does not saturate even at the maximum applied field of 7 T. The parent compound Mn$_{1.4}$PtSn bears a moment of 4.6 $\mu_B$/f.u. at 7 T at 2 K. For both, Pd and Rh substitution, the magnetization at 7 T and 2 K decreases with increasing degree of substitution *x* or *y*. However, the variation is rather small in the case of Pd (*x*) as compared to Rh (*y*) substitution even for the same amount of substitution. For instance, the magnetizations at 7 T approach to 4.24 and 3.85 $\mu_B$/f.u. for $x = 0.3$ and $y = 0.3$, respectively. An approximately linear decrease of the magnetization is observed with increasing *y*, *i.e.* decreasing number of valence electrons, for Rh contents up to $y = 0.5$, which demonstrates a Slater-Pauling-like reduction [42].

### D. Topological Hall effect

In Fig. 5(a) we show the total Hall resistivity together with the magnetization as a function of an external field for Mn$_{1.4}$Pt$_{0.9}$Pd$_{0.1}$Sn. Two distinctions in the Hall resistivity $\rho_{yx}(H)$ as compared to the magnetization $M(H)$ are observed: (1) An enhanced feature just below the field, above which the magnetization tends to saturate and (2) A large coercivity (~

−450 Oe) in the Hall resistivity hysteresis (within ±0.5 T) which is reversed to that of the magnetic isotherm. As mentioned in the experimental section the maximum error in the magnetic field in our superconducting magnets is 20 Oe, such small field errors cannot be responsible for the hysteresis loop reversals observed for the present material. Indeed, the observation of enhanced feature below the saturation field in $\rho_{yx}(H)$ indicates that actually a non-coplanar spin configuration emerges from the canted spins below $T_{SR}$ which is induced by application of a magnetic field in the Hall experiment and gives rise to a THE. On the other hand, the inverse hysteretic observation at lower field in $\rho_{yx}(H)$ suggests the presence of aSKs whose cores are antiparallel to the field direction and which are stabilized by the MCA. In B20 compounds this effect is observed when the anisotropy is induced by the thin film limit [7,22,43], while in the compounds with $D_{2d}$ symmetry the anisotropy is inherent to the crystal lattice. In addition, microstructural defects like twins and grain boundaries of the polycrystalline samples may also act as pinning centers to stabilize the aSKs. Previous reports on analogous Heusler compounds *e.g.* bulk $Mn_2PtSn$ [44] and thin films of $Mn_2PtSn$ [45], $Mn_{2-x}PtSn$ [46] and $Mn_2RhSn$ [47] also revealed a THE but did not show an inverse hysteresis in the $\rho_{yx}$ loop. In Fig. 5(b) the topological Hall resistivity, $\rho^T$ at $T = 2$, 120 and 150 K is shown (the method for the extraction of $\rho^T$ is shown in Supplemental Material [38]). $\rho^T$ displays an inverse hysteresis as compared to the magnetization at 2 and 120 K (< $T_{SR}$), whereas no hysteresis is present at 150 K (> $T_{SR}$). In Fig. 5(c) we plot three features of the THE as a function of temperature: $\rho^T_{max}$ the average of the maximum values with the sign determined by the positive field value, $\rho^T_0 = (\rho^T_{0-} - \rho^T_{0+})/2$ the zero field value, and $H^T_C = (H^+_C - H^-_C)/2$ the coercivity in the $\rho^T$ loops. The −(+) sign corresponds to a decreasing (increasing) field sweep. $\rho^T_{max}$ reaches a maximum of 0.42 μΩ cm at 100 K and decreases to zero above $T_{SR}$. It is unclear whether the finite but small $\rho^T_{max}$ value at 150 K ($T > T_{SR}$) in the collinear magnetic phase is an artifact or actually is of topological origin. However, at higher temperatures ($T >> T_{SR}$) $\rho^T_{max}$ is nearly zero (see Fig. S11 of the Supplemental Material [38]). $H^T_C$ (red) continuously decreases with temperature, due to the fluctuation of the moments at higher temperatures. $\rho^T_0$, most interestingly, is constant, −0.08 μΩ cm, up to $T_{SR}$, a further indication of an aSK state, where $\rho^T$ is expected to be independent of the variation in longitudinal resistivity [19]. Further evidence of aSKs at $H = 0$ is obtained by the opposite signs of $\rho^T_{max}$ and $\rho^T_0$, which can only be caused by a sign change in the topological texture on going from a NCP structure to an aSK phase. In the work of Nayak *et al.* [15], the aSK phase was even stable above $T_{SR}$ in thin lamellae of $Mn_{1.4}Pt_{0.9}Pd_{0.1}Sn$, which is probably related to the reduced dimensionality. By contrast, the present study of Hall experiments demonstrates that bulk polycrystalline materials show aSKs only in the non-collinear magnetic phase ($< T_{SR}$).

Upon increasing the Pd content up to $x = 0.3$, the largest possible Pd substitution level, only slight changes in magnetic and transport properties are observed (see the Supplemental Material [38]). This implies that a moderate change in SOC due to variation of the Pd substitution does not change much the physical situation reflected by Fig. 5. By contrast, substitution of Rh for Pt has more pronounced implications for the properties. In particular, the magnetization features a strong Slater-Pauling-like reduction with decreasing electron count as well as a decrease in $T_C$ (see Fig. 4). Analysis of the field dependence of the magnetization indicates that the MCA in the present system is strengthened by increasing the Rh content (see Fig. S6 of Supplemental Material [38]).

In Fig. 6 the Hall transport properties of $Mn_{1.4}Pt_{1-y}Rh_ySn$ are summarized. In Fig. 6(a) we show the experimental Hall resistivity corrected for the normal Hall contribution $\rho_{yx} - R_0\mu_0H (= \rho^A + \rho^T)$ for decreasing and increasing field sweeps at 2 K for $0.1 \leq y \leq 0.6$. Here $R_0$ is the normal Hall coefficient. The calculated anomalous Hall resistivity $\rho^A (= b\rho_{xx}^2 M)$ is also shown. The compounds for $y < 0.3$ display similar features in the Hall signals as observed for the $Mn_{1.4}Pt_{1-x}Pd_xSn$ compounds. However, further increment in the Rh content results in a larger coercivity in the magnetization, $H_C^M$ and the Hall resistivity, $H_C^H$ follows the magnetization hysteresis with respect to the sweep directions. In addition, the magnitude of the coercivity of $\rho_{yx}$ differs significantly from $H_C^M$ up to $y = 0.5$, whereas it is nearly equal for $y \geq 0.6$ (see Table 1). This is further shown as a schematic of arrows in Fig. 6(a). The increase in the magnetization coercivity ($H_C^M$) with increase in the Rh content is consistent with the increasing MCA and anisotropy field estimated from the law of approach to saturation (see Fig. S6 [38]).

In Fig. 6(b) and (c) we show the $\rho^T$ curves at 2 K and the temperature dependence of the topological Hall parameters ($\rho^T$ and $H_C^T$ values), respectively. A sign change from negative to positive in $H_C^T$ and consequently also in $\rho_0^T$ occurs at $y = 0.3$. For $y = 0.5$, the topological parameters reach a maximum. As indicated by the neutron diffraction studies, up to $y = 0.5$ a non-collinear spin structure below $T_{SR}$ is retained, with a nearly constant angle between the two Mn sublattices of ~ 80°. However, for $y \geq 0.6$, the Mn sublattices approach the collinear FiM state in the whole temperature range and the topological Hall parameters sharply decrease.

In Fig. 7 the results of the topological ($H_C^T$) and magnetic ($H_C^M$) coercive field are presented together with $\rho_{max}^T$ and $\rho_0^T$ at 2 K. The dependence of the parameters on the Rh content can be related to the presence of three distinct magnetic structures. For Rh contents 0.6 ≤ y ≤ 0.8 (yellow background) the spins align antiferromagnetically with a larger moment on the Mn1 site in a FiM configuration. Here, $H_C^T$ is significantly low owing to the similar coercivity in $\rho_{yx}(H)$ and $M(H)$ curves. Nevertheless, we found a finite negative $\rho_{max}^T$ and $\rho_0^T$ which may be due to the small canting of the antiferromagnetic order vector or it can be an error generated in the extraction of the topological Hall resistivity (see the Supplemental Material [38]). Both $\rho_{max}^T$ and $\rho_0^T$ monotonously decrease to zero as the Rh content reaches y = 0.8.

In the range of 0.25 < y < 0.6 (green background), the spins form a NCP spin structure induced by the magnetic field. The variation of the MCA and the DMI with the SOC (coupling constant $\lambda$) and magnetization ($M$) can be expressed by the relations $K_{ani} \sim \lambda^2$ [48] and $D \sim \lambda M^2$ [49], where $K_{ani}$ and $D$ are the anisotropy and DMI constant, respectively. Moreover, their strengths largely depend on the electronic states nearby the Fermi level which are expected to be varied to a large extent by Rh substitution (changing the number of valence electrons). A lower magnetization in this composition range may impede the formation of an aSK state as a result of overcompensation of the DMI by the MCA. Both, $\rho_{max}^T$ and $\rho_0^T$ are additionally maximized in this region, however with an opposite sign at the maximum (y = 0.5). This is due to the change in the electronic structure, as also evidenced in the normal Hall effect, where the slope of $\rho_{yx}$ changes sign near y = 0.5 (see Fig. S13 [38]).

Finally, the aSK phase (blue background) occurs in both Rh and Pd substituted samples in the composition ranges 0 < y < 0.25 and 0 ≤ x ≤ 0.3, respectively. The aSK phase is stabilized at lower fields ($\mu_0 H \leq |0.5|$ T) and remained stable at zero field. The $\rho_{max}^T$ has a nearly constant value of ~ 0.2 $\mu\Omega$ cm. Furthermore, $H_C^M$ is smallest in this range but its sign is also opposite to that of $H_C^T$. This is a key feature of the formation of aSKs. In this composition range, due to the enhanced magnetization, the DMI is sufficiently large to induce transformation of the NCP structure to the aSK phase, whereas the MCA is primarily important for the survival of the aSK at zero field. The opposite sign for $\rho_0^T$ and $\rho_{max}^T$ is due to the change in topology of the magnetic spin structure.

The $\rho_{max}^T$ found in the present system is due to the NCP spin structure, which displays the largest value of ~ 0.5 $\mu\Omega$ cm, and is comparable to that observed in other similar Heusler compounds as bulk $Mn_2PtSn$ (1.53 $\mu\Omega$ cm) [44], thin films of $Mn_2PtSn$ (0.5 $\mu\Omega$ cm) [45] and Mn$_{2-x}$PtSn (1.2 μΩ cm) [46]. However, these cases did not show the reversal of hysteresis in the Hall resistivity compared to that of the magnetization, which is attributed to the aSK phase. The presence of aSKs results in a finite value of $\rho_0^T$ with a negative sign, which has the largest magnitude of ~ 0.08 μΩ cm for $x = 0.1$. The THE in the B20 compounds due to the stabilization of skyrmions has been well described by the size of their spin texture. The emergent magnetic field ($B_{em}$) shows quadratic decrement with the size ($L_{sk}$) of the skyrmion in the adiabatic limit, $B_{em} \propto 1/L_{sk}^2$ [18]. For instance, MnGe [19] and FeGe [22] display $\rho^T$ values of 0.16 and 0.08 μΩ cm at a finite field for $L_{sk}$ = 3 and 70 nm, respectively. Intriguingly, despite the comparatively large size of the aSKs in Mn$_{1.4}$Pt$_{0.9}$Pd$_{0.1}$Sn (~150 nm) [15], we find a large topological Hall resistivity even at zero fields. This is likely due to the role of the electronic structure in the topological Hall constant $R^T = \rho^T/B_{em}$ [24], and provides the possibility to engineer even a larger THE in aSK phases with smaller periods.

## IV. CONCLUSIONS

We have shown the THE due to the robust formation of antiskyrmions below $T_{SR}$ in the bulk Mn$_{1.4}$PtSn system with fractional substitution of Pt by Pd and Rh. The competition between the DMI and the MCA is crucial for the transformation of a non-coplanar spin structure to the antiskyrmion phase. Furthermore, we expect the antiskyrmion phase to be stable for wider composition and temperature ranges in thin plate samples, as indicated by our previous work [Nature 548, 561 (2017)]. This leads to the possibility of having smaller sized antiskyrmions, which can also be achieved by substitution of Pt by another heavy metal like Ir. Lastly, the ferrimagnetic configurations found for large Rh substitutions open the door to antiferromagnetic spintronics based on aSKs. Particularly useful would be ferrimagnetic materials tuned to the compensation point as compensated ferrimagnetic aSKs would travel in a straight line trajectory and yet would be easily detectable due to different electronic properties of the constituent elements [54].

## ACKNOWLEDGEMENT

The authors acknowledge funding by the Deutsche Forschungsgemeinschaft (DFG, German Research Foundation) under SPP 2137 (Project No. 403502666) and EU FET Open RIA Grant No. 766566 "ASPIN". We would like to thank Ajaya K. Nayak and Walter Schnelle for discussion.

* claudia.felser@cpfs.mpg.de

**REFERENCES**

[1] S. Parkin and S.-H. Yang, Nat. Nanotechnol. **10**, 195 (2015).

[2] A. Fert, V. Cros, and J. Sampaio, Nat. Nanotechnol. **8**, 152 (2013).

[3] D. Pinna, F. A. Araujo, J.-V. Kim, V. Cros, D. Querlioz, P. Bessiere, J. Droulez, and J. Grollier, Phys. Rev. Appl. **9**, 064018 (2018).

[4] Z. He and D. Fan, ArXiv:1705.02995v1.

[5] X. Z. Yu, N. Kanazawa, W. Z. Zhang, T. Nagai, T. Hara, K. Kimoto, Y. Matsui, Y. Onose, and Y. Tokura, Nat. Commun. **3**, 988 (2012).

[6] F. Jonietz, S. Mühlbauer, C. Pfleiderer, A. Neubauer, W. Münzer, A. Bauer, T. Adams, R. Georgii, P. Böni, R. A. Duine, K. Everschor, M. Garst, and A. Rosch, Science **330**, 1648 (2010).

[7] J. C. Gallagher, K. Y. Meng, J. T. Brangham, H. L. Wang, B. D. Esser, D. W. McComb, and F. Y. Yang, Phys. Rev. Lett. **118**, 027201 (2017).

[8] A. Neubauer, C. Pfleiderer, B. Binz, A. Rosch, R. Ritz, P. G. Niklowitz, and P. Böni, Phys. Rev. Lett. **102**, 186602 (2009).

[9] D. Xiao, M.-C. Chang, and Q. Niu, Rev. Mod. Phys. **82**, 1959 (2010).

[10] K. Manna, Y. Sun, L. Muechler, J. Kübler, and C. Felser, Nat. Mater. Rev. **3**, 244 (2018).

[11] A. N. Bogdanov and D. A. Yablonskiĭ, Sov. Phys. JETP **68**, 101 (1989).

[12] U. K. Rößler, A. N. Bogdanov, and C. Pfleiderer, Nature **442**, 797 (2006).

[13] S. Mühlbauer, B. Binz, F. Jonietz, C. Pfleiderer, A. Rosch, A. Neubauer, R. Georgii, and P. Böni, Science **323**, 915 (2009).

[14] A. N. Bogdanov, U. K. Rößler, M. Wolf, and K.-H. Müller, Phys. Rev. B **66**, 214410 (2002).

[15] A. K. Nayak, V. Kumar, T. Ma, P. Werner, E. Pippel, R. Sahoo, F. Damay, U. K. Rößler, C. Felser, and S. S. P. Parkin, Nature **548**, 561 (2017).

[16] S. Huang, C. Zhou, G. Chen, H. Shen, A. K. Schmid, K. Liu, and Y. Wu, Phys. Rev. B **96**, 144412 (2017).

[17] W. Koshibae and N. Nagaosa, Nat. Commun. **7**, 10542 (2016).

[18] N. Nagaosa and Y. Tokura, Nat. Nanotechnol. **8**, 899 (2013).

[19] N. Kanazawa, Y. Onose, T. Arima, D. Okuyama, K. Ohoyama, S. Wakimoto, K. Kakurai, S. Ishiwata, and Y. Tokura, Phys. Rev. Lett. **106**, 156603 (2011).

[20] M. Onoda, G. Tatara, and N. Nagaosa, J. Phys. Soc. Japan **73**, 2624 (2004).

[21] T. Yokouchi, N. Kanazawa, A. Tsukazaki, Y. Kozuka, M. Kawasaki, M. Ichikawa, F. Kagawa, and Y. Tokura, Phys. Rev. B **89**, 064416 (2014).

[22] S. X. Huang and C. L. Chien, Phys. Rev. Lett. **108**, 267201 (2012).

[23] Y. Li, N. Kanazawa, X. Z. Yu, A. Tsukazaki, M. Kawasaki, M. Ichikawa, X. F. Jin, F. Kagawa, and Y. Tokura, Phys. Rev. Lett. **110**, 117202 (2013).

[24] C. Franz, F. Freimuth, A. Bauer, R. Ritz, C. Schnarr, C. Duvinage, T. Adams, S. Blügel, A. Rosch, Y. Mokrousov, and C. Pfleiderer, Phys. Rev. Lett. **112**, 186601 (2014).

[25] Y. T. Y. Taguchi, Y. Oohara, H. Yoshizawa, N. Nagaosa, and Y. Tokura, Science **291**, 2573 (2001).

[26] Y. MacHida, S. Nakatsuji, Y. Maeno, T. Tayama, T. Sakakibara, and S. Onoda, Phys. Rev. Lett. **98**, 057203 (2007).

[27] C. Sürgers, G. Fischer, P. Winkel, and H. V. Löhneysen, Nat. Commun. **5**, 3400 (2014).

[28] B. G. Ueland, C. F. Miclea, Y. Kato, A. Valenzuela, R. D. McDonald, R. Okazaki, P. H. Tobash, M. A. Torrez, F. Ronnig, R. Movshovich, Z. Fisk, E. D. Bauer, I. Martin, and J. D. Thompson, Nat. Commun. **3**, 1067 (2012).

[29] F. W. Fabris, P. Pureur, J. Schaf, V. N. Vieira, and I. A. Campbell, Phys. Rev. B **74**, 214201 (2006).

[30] T. Taniguchi, K. Yamanaka, H. Sumioka, T. Yamazaki, Y. Tabata, and S. Kawarazaki, Phys. Rev. Lett. **93**, 246605 (2004).

[31] Y. Ohuchi, Y. Kozuka, M. Uchida, K. Ueno, A. Tsukazaki, and M. Kawasaki, Phys. Rev. B **91**, 245115 (2015).

[32] L. Vistoli, W. Wang, A. Sander, Q. Zhu, B. Casals, R. Cichelero, A. Barthélémy, S. Fusil, G. Herranz, S. Valencia, R. Abrudan, E. Weschke, K. Nakazawa, H. Kohno, J.

Santamaria, W. Wu, V. Garcia, and M. Bibes, Nat. Phys. **15**, 67 (2019).

[33] O. Meshcheriakova, S. Chadov, A. K. Nayak, U. K. Rößler, J. Kübler, G. André, A. A. Tsirlin, J. Kiss, S. Hausdorf, A. Kalache, W. Schnelle, M. Nicklas, and C. Felser, Phys. Rev. Lett. **113**, 087203 (2014).

[34] J. Rodríguez-Carvajal, Phys. B **192**, 55 (1993).

[35] V. F. Sears, *International Tables for Crystallography*, C (Kluwer Academic Publishers, Dordrecht/Boston/London, 1995) **C 383**.

[36] P. J. Brown, *International Tables for Crystallography*, C (Kluwer Academic Publishers, Dordrecht/Boston/London, 1995) **C 391**.

[37] G. Mastrogiacomo, J. F. Löffler, and N. R. Dilley, Appl. Phys. Lett. **92**, 082501 (2008). Application Note 1070-207, Using PPMS Superconducting Magnets at Low Fields, Quantum Design, San Diego (2009). Application Note 1500-021, Correcting for the Absolute Field Error using the Pd Standard, San Diego (2017).

[38] See Supplemental Material at http://link.aps.org/supplemental/... for further details about data evaluation and additional structural, neutron, magnetic characterization, and transport data, which includes ref. [50-53].

[39] L. Wollmann, S. Chadov, J. Kübler, and C. Felser, Phys. Rev. B **90**, 214420 (2014).

[40] L. Wollmann, S. Chadov, J. Kübler, and C. Felser, Phys. Rev. B **92**, 064417 (2015).

[41] P. Vir, J. Gayles, A. S. Sukhanov, N. Kumar, F. Damay, Y. Sun, J. Kübler, C. Shekhar, and C. Felser, Phys. Rev. B **99**, 140406(R) (2019).

[42] S. Chadov, J. Kiss, and C. Felser, Adv. Funct. Mater. **23**, 832 (2013).

[43] N. Kanazawa, M. Kubota, A. Tsukazaki, Y. Kozuka, K. S. Takahashi, M. Kawasaki, M. Ichikawa, F. Kagawa, and Y. Tokura, Phys. Rev. B **91**, 041122(R) (2015).

[44] Z. H. Liu, A. Burigu, Y. J. Zhang, H. M. Jafri, X. Q. Ma, E. K. Liu, W. H. Wang, and G. H. Wu, Scr. Mater. **143**, 122 (2018).

[45] Y. Li, B. Ding, X. Wang, H. Zhang, W. Wang, and Z. Liu, Appl. Phys. Lett. **113**, 062406 (2018).

[46] P. Swekis, A. Markou, D. Kriegner, J. Gayles, R. Schlitz, W. Schnelle, S. T. B. Goennenwein, and C. Felser, Phys. Rev. Mater. **3**, 013001(R) (2019).

[47] K. G. Rana, O. Meshcheriakova, J. Kübler, B. Ernst, J. Karel, R. Hillebrand, E. Pippel, P. Werner, A. K. Nayak, C. Felser, and S. S. P. Parkin, New J. Phys. **18**, 085007 (2016).

[48] P. Bruno, Phys. Rev. B **39**, 865 (1989).

[49] F. Freimuth, R. Bamler, Y. Mokrousov, and A. Rosch, Phys. Rev. B **88**, 214409 (2013).

[50] Y. Tokunaga, X. Z. Yu, J. S. White, H. M. Rønnow, D. Morikawa, Y. Taguchi, and Y. Tokura, Nat. Commun. **6**, 7638 (2015).

[51] K. Karube, J. S. White, N. Reynolds, J. L. Gavilano, H. Oike, A. Kikkawa, F. Kagawa, Y. Tokunaga, H. M. Rønnow, Y. Tokura, and Y. Taguchi, Nat. Mater. **15**, 1237 (2016).

[52] S. V. Andreev, M. I. Bartashevich, V. I. Pushkarskya, V. N. Maltsev, L. A. Pamyatnykh, E. N. Tarasov, N. V. Kudrevatykh, and T. Goto, J. Alloys Compd. **260**, 196 (1997).

[53] Y. Huh, P. Kharel, A. Nelson, V. R. Shah, J. Pereiro, P. Manchanda, A. Kashyap, R. Skomski, and D. J. Sellmyer, J. Phys. Condens. Matter **27**, 076002 (2015).

[54] L. Caretta, M. Mann, F. Büttner, K. Ueda, B. Pfau, C. M. Günther, P. Hessing, A. Churikova, C. Klose, M. Schneider, D. Engel, C. Marcus, D. Bono, K. Bagschik, S. Eisebitt, and G. S. D. Beach, Nat. Nanotechnol. **13**, 1154 (2018).

**FIGURES and TABLE**

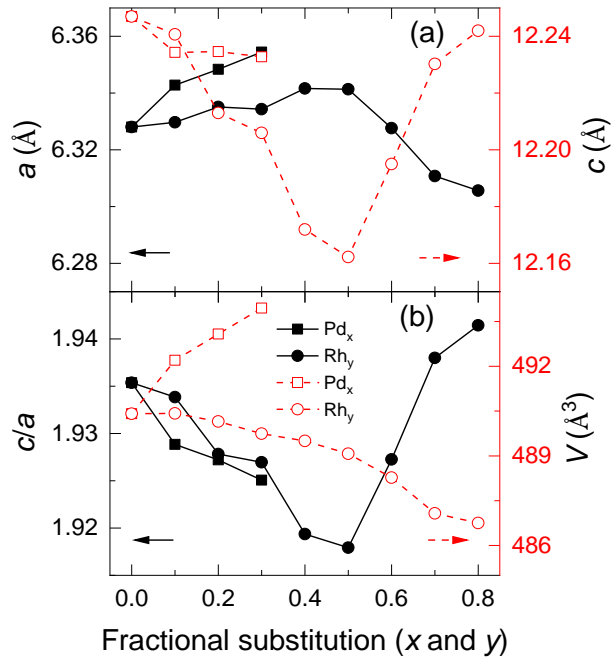

FIG. 1. Lattice constants, *a* and *c*, *c*/*a* ratio and cell volume *V* of Mn$_{1.4}$Pt$_{1-x}$Pd$_x$Sn and Mn$_{1.4}$Pt$_{1-y}$Rh$_y$Sn. The size of error bars is less than the symbol size.

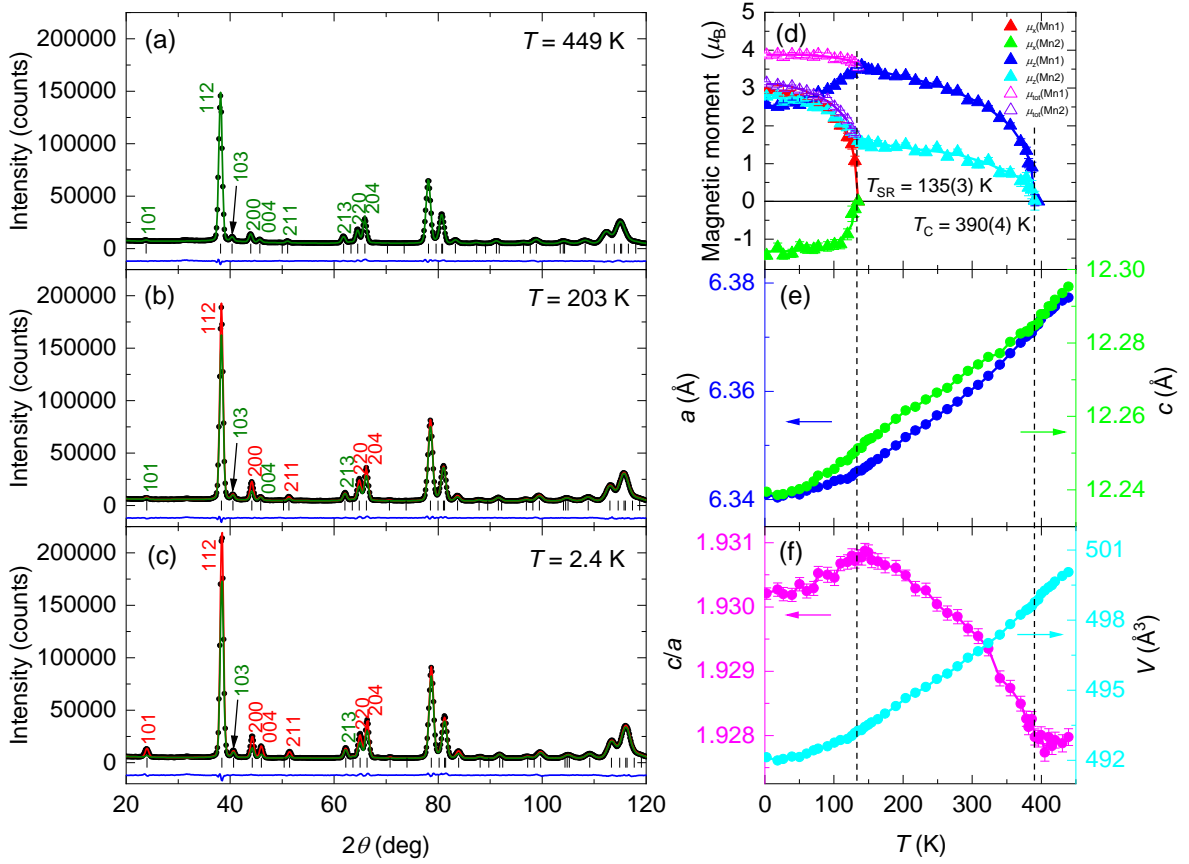

FIG. 2. Results of the Rietveld refinements of the neutron powder diffraction data of Mn$_{1.4}$Pt$_{0.9}$Pd$_{0.1}$Sn collected with $\lambda$ = 2.43 Å at (a) 449, (b) 203 and (c) 2.4 K, respectively. The crystal structure was refined in the tetragonal space group $I\bar{4}2d$. At 203 K the strongest magnetic intensity appears at the positions of the reflections 112, 200, 211, 220, and 204 (marked in red) indicating a ferromagnetic spin alignment of the Mn atoms. At 2.4 K the reflections 101 and 004 show additional magnetic intensities indicating a spin reorientation transition. The calculated patterns (crystal structure in green, crystal and magnetic structure in red) are compared with the observed ones (black circles). The difference patterns (blue) as well as the positions (black bars) of the Bragg reflections are shown. In (d), (e), and (f) the temperature dependence of the total magnetic moment $\mu_{tot}$ and the magnetic moment components $\mu_x$ and $\mu_z$ of the Mn atoms, the lattice parameters $a$ and $c$, and the $c/a$ ratio and the unit cell volume $V$ in Mn$_{1.4}$Pt$_{0.9}$Pd$_{0.1}$Sn are depicted, respectively. The size of error bars of $a$, $c$ and $V$ is less than the symbol size. Ferromagnetic ordering sets in at the Curie temperature $T_C$ = 390(4) K. Below $T_{SR}$ = 135(3) K the $x$ components of the Mn1 and Mn2 moments show a ferrimagnetic spin alignment. The dotted vertical lines are a guide for the eye.

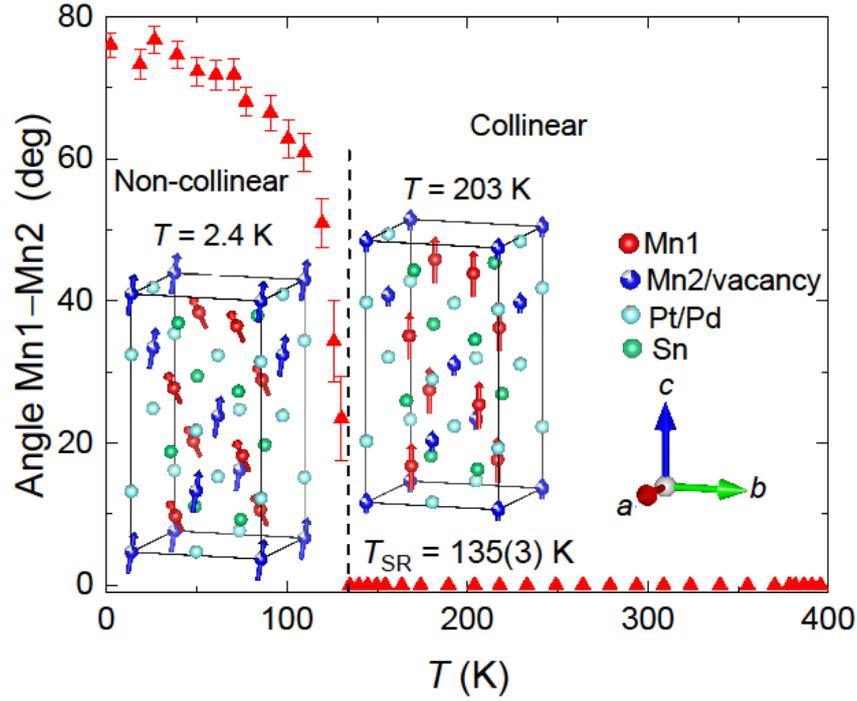

FIG. 3. Magnetic structure of $Mn_{1.4}Pt_{0.9}Pd_{0.1}Sn$ at 2.4 and 203 K, and the variation of the angle between Mn1 and Mn2 moments as a function of temperature.

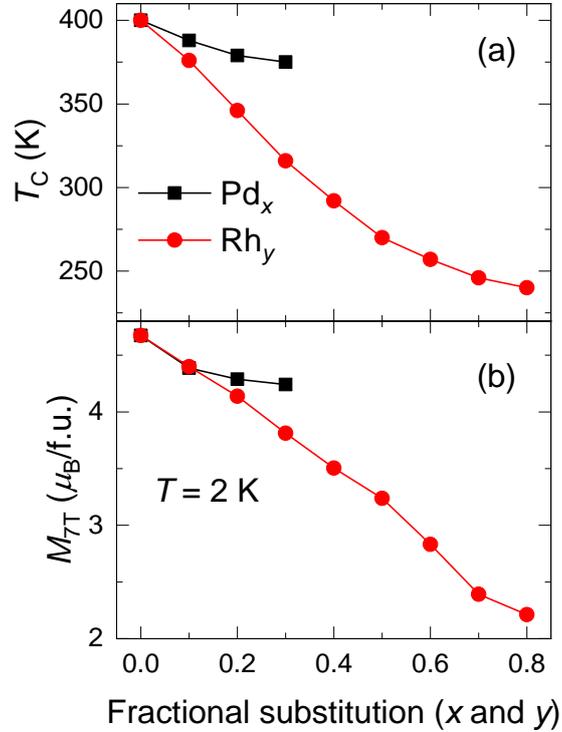

FIG. 4. (a) Curie temperature and (b) magnetization at 7 T and 2 K of $Mn_{1.4}Pt_{1-x}Pd_xSn$ and $Mn_{1.4}Pt_{1-y}Rh_ySn$.

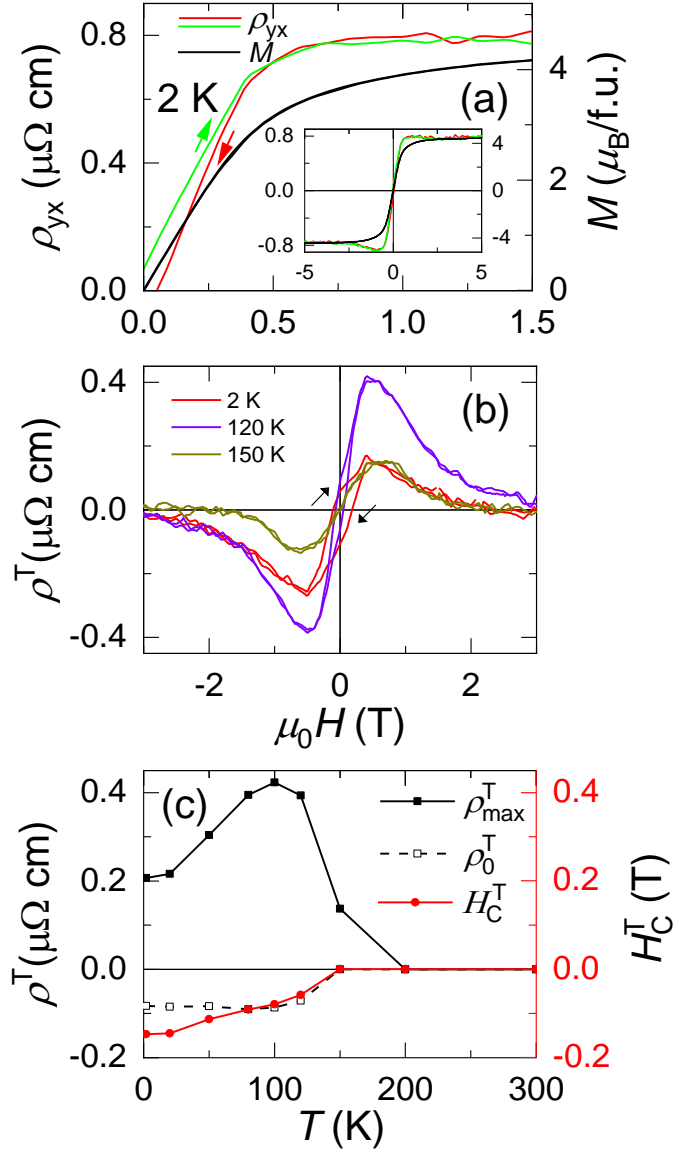

FIG. 5. (a) Hall resistivity (left axis) and magnetization (right axis) of $Mn_{1.4}Pt_{0.9}Pd_{0.1}Sn$ between 0 and 1.5 T. Decreasing and increasing field sweep directions in $\rho_{yx}$ (red and green lines, respectively) are indicated by arrows. The inset shows full hysteresis loops between −5 and +5 T. (b) Topological Hall resistivity at 2, 120 and 150 K. (c) Temperature dependence of maximum and zero field value and coercivity of the topological Hall resistivity.

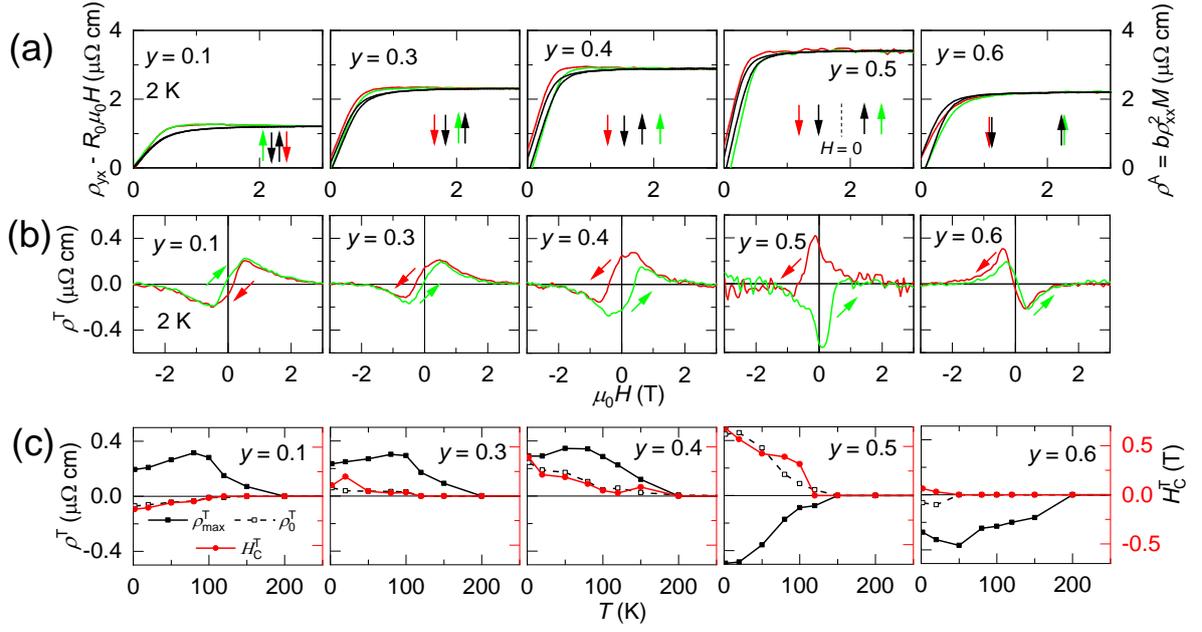

FIG. 6. (a) Corrected Hall resistivity $\rho_{yx} - R_0\mu_0 H$ (left axis, red and green lines) and the calculated anomalous Hall resistivity $\rho^A (= b\rho_{xx}^2 M)$ (right axis, black lines) at 2 K of $Mn_{1.4}Pt_{1-y}Rh_y Sn$. The black arrows detail the magnetization hysteresis around zero fields with the down (up) arrow for the decreasing (increasing) field sweep. The red (green) arrow shows the position of the Hall resistivity for decreasing (increasing) field sweep around zero fields with respect to the magnetization. (b) Topological Hall resistivity at 2 K and (c) the temperature dependent parameters characterizing $\rho^T$.
19

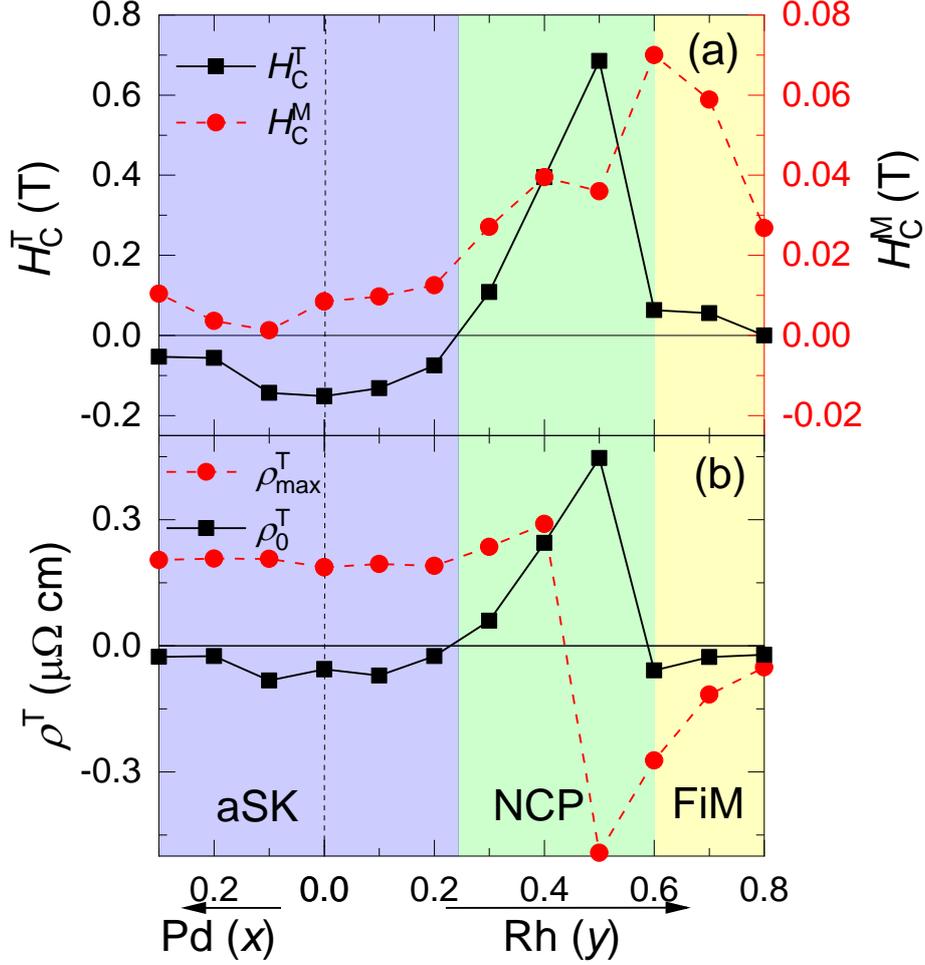

FIG. 7. (a) Coercive field of topological Hall resistivity (left axis) and magnetization (right axis) and (b) topological Hall resistivity of $Mn_{1.4}Pt_{1-x}Pd_xSn$ and $Mn_{1.4}Pt_{1-y}Rh_ySn$ at 2 K. The compounds are classified in three composition ranges corresponding to the aSK phase, NCP spin structure and FiM configuration which are indicated by blue, green and yellow colors.

TABLE 1: Coercive fields (in Oe) of $Mn_{1.4}Pt_{1-y}Rh_ySn$.

| $y$ | Coercivity of $M(H)$, $H_C^M$ | Coercivity of $\rho_{yx}$, $H_C^H$ | Coercivity of $\rho^T$, $H_C^T$ |
| --- | --- | --- | --- |
| 0.1 | 96 ± 3 | -169 ± 31 | -1357 ± 43 |
| 0.3 | 252 ± 17 | 285 ± 77 | 1105 ± 21 |
| 0.4 | 418 ± 16 | 690 ± 56 | 4194 ± 110 |
| 0.5 | 396 ± 49 | 1030 ± 91 | 5802 ± 460 |
| 0.6 | 672 ± 35 | 769 ± 66 | 633 ± 1 |

**Content**

**I. Powder x-ray diffraction (XRD)**

FIG. S1. Room temperature powder x-ray diffraction patterns of $Mn_{1.4}Pt_{1-x}Pd_xSn$ and $Mn_{1.4}Pt_{1-y}Rh_ySn$.

**II. Neutron powder diffraction**

**Refinement results for $Mn_{1.4}Pt_{0.9}Pd_{0.1}Sn$**

TABLE S1. Results of the neutron diffraction study of $Mn_{1.4}Pt_{0.9}Pd_{0.1}Sn$.

**Comparison of magnetic moments derived from neutron and magnetization data**

**Symmetry analysis**

**Neutron studies on $Mn_{1.4}Pt_{1-y}Rh_ySn$**

FIG. S2. Rietveld refinement results for $Mn_{1.4}Pt_{1-y}Rh_ySn$ with (a) $y = 0.4$ and (b) $y = 0.6$ at 1.5 K.

**III. Magnetic characterization**

FIG. S3. Magnetization measurements of $Mn_{1.4}Pt_{0.9}Pd_{0.1}Sn$.

FIG. S4. Temperature dependent real part $\chi'$ of the AC susceptibility and magnetization $M(T)$ for $Mn_{1.4}Pt_{1-x}Pd_xSn$ and $Mn_{1.4}Pt_{1-y}Rh_ySn$.

FIG. S5. Magnetic isotherms of $Mn_{1.4}Pt_{1-y}Rh_ySn$.

FIG. S6. Calculated magnetocrystalline anisotropy constant $K_1$ and anisotropy field $H_K$ at 200 K of $Mn_{1.4}Pt_{1-x}Pd_xSn$ and $Mn_{1.4}Pt_{1-y}Rh_ySn$.

**IV. Transport measurements**

FIG. S7. Transport properties of $Mn_{1.4}Pt_{1-x}Pd_xSn$ and $Mn_{1.4}Pt_{1-y}Rh_ySn$.

FIG. S8 and S9. Method of extraction of topological Hall resistivity.

FIG. S10 log-log plot of $\rho_{xx}$ vs $\rho^A$.

FIG. S11. Topological Hall resistivity above $T_{SR}$ of Mn$_{1.4}$Pt$_{0.9}$Pd$_{0.1}$Sn.

FIG. S12. Hall resistivity data of Mn$_{1.4}$Pt$_{1-y}$Rh$_y$Sn.

FIG. S13. The experimental Hall resistivity of Mn$_{1.4}$Pt$_{1-y}$Rh$_y$Sn.

FIG. S14. Topological Hall resistivity of Mn$_{1.4}$Pt$_{1-y}$Rh$_y$Sn ($y = 0.7$ and 0.8).

FIG. S15. Anomalous Hall resistivity and conductivity of Mn$_{1.4}$Pt$_{1-x}$Pd$_x$Sn and Mn$_{1.4}$Pt$_{1-y}$Rh$_y$Sn.

## I. Powder x-ray diffraction (XRD)

We show room temperature XRD patterns in Fig. S1.

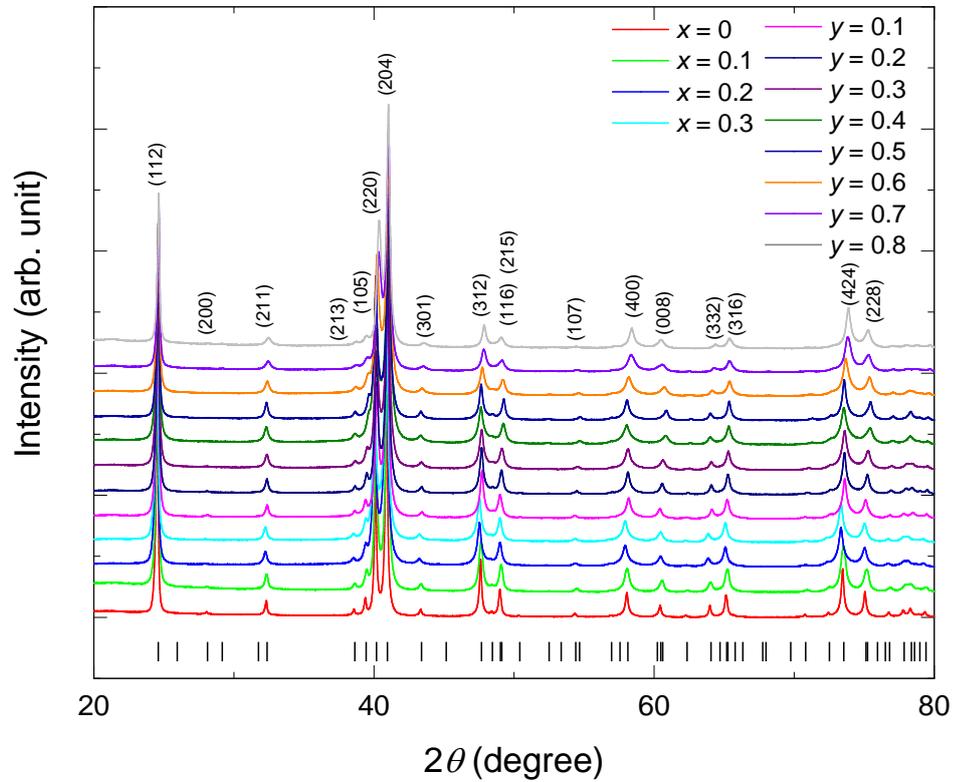

FIG. S1. Room temperature powder x-ray diffraction patterns of $Mn_{1.4}Pt_{1-x}Pd_xSn$ and $Mn_{1.4}Pt_{1-y}Rh_ySn$. Vertical bars indicate the calculated positions of the Bragg reflections for $x = 0$. The lattice parameters obtained by Rietveld refinement of the patterns are shown in Fig. 1.

## II. Neutron powder diffraction

**Refinement results for Mn$_{1.4}$Pt$_{0.9}$Pd$_{0.1}$Sn**

TABLE S1. Results of the neutron diffraction study of Mn$_{1.4}$Pt$_{0.9}$Pd$_{0.1}$Sn. The residual for the crystal and magnetic structure are defined as $R_F = \sum||F_{obs}| - |F_{calc}||/\sum|F_{obs}|)$ and $R_M = \sum||I_{obs}| - |I_{calc}||/\sum|I_{obs}|)$, respectively.

| Mn$_{1.4}$Pt$_{0.9}$Pd$_{0.1}$Sn | at 2.4 K | at 203 K | at 449 K |
| --- | --- | --- | --- |
| $a$ [Å] | 6.34100(18) | 6.35154(17) | 6.37744(19) |
| $c$ [Å] | 12.2395(4) | 12.2616(4) | 12.2999(5) |
| $c/a$ | 1.9302(13) | 1.9305(12) | 1.9287(13) |
| $V$ [Å$^3$] | 492.13(3) | 494.66(5) | 500.26(4) |
| $x$(Mn1) | 0.7323(6) | 0.7313(5) | 0.7309(8) |
| $z$(Pt/Pd) | 0.01419(14) | 0.01398(10) | 0.01393(16) |
| $x$(Sn) | 0.2774(4) | 0.2772(4) | 0.2764(4) |
| $occ$(Mn2) | 0.772 | 0.772 | 0.772(19) |
| $occ$(Pt/Pd) | 0.902/0.089 | 0.902/0.089 | 0.902(22)/0.089(22) |
| $d$(Mn1-Mn2) | 2.7813(25) | 2.7900(19) | 2.8019(29) |
| $R_F$ | 0.0158 | 0.0157 | 0.0168 |
| $\mu_x$(Mn1) | 2.89(7) | - | - |
| $\mu_z$(Mn1) | 2.56(6) | 3.32(3) | - |
| $\mu_{tot}$(Mn1) | 3.87(5) | 3.32(3) | - |
| $\mu_x$(Mn2) | −1.65(11) | - | - |
| $\mu_z$(Mn2) | 3.13(7) | 1.70(6) | - |
| $\mu_{tot}$(Mn2) | 3.54(5) | 1.70(6) | - |
| $R_M$ | 0.0162 | 0.0153 | - |

**Comparison of magnetic moments derived from neutron and magnetization data**

At 203 K the neutron data give a full magnetic moment per formula unit (f.u.) $\mu_{s,z} = \mu_z$(Mn1) + 0.386 × $\mu_z$(Mn2) = 3.32 + 0.386 × 1.70 $\mu_B$ = 3.98 $\mu_B$/f.u. Here, the factor 0.386 takes into account the multiplicity and the partial occupancy of the Mn2 site. The total moment matches well with that from the magnetization measurements $\mu_{s,M}$ = 4.01 $\mu_B$/f.u. (at 7 T) obtained at 200 K. At 2 K the magnetization reaches $\mu_{s,M}$ = 4.43 $\mu_B$/f.u. (at 7 T). At 2.4 K the neutron data gives $\mu_{s,z} = \mu_z(Mn1) + 0.386 \mu_z(Mn2) = 2.56 + 0.386 \times 3.13 \mu_B = 3.77 \mu_B$/f.u. for the $z$ component. For the $x$ component one obtains $\mu_{s,x} = \mu_x(Mn1) + 0.386 \mu_x(Mn2) = 2.89 - 0.386 \times 1.65 \mu_B = 2.26 \mu_B$/f.u. This gives in total $\mu_{s,tot} = 4.40 \mu_B$/f.u. [$\mu_{s,tot} = \sqrt{(\mu_{s,x}^2 + \mu_{s,z}^2)}$], which again matches well with the value from the magnetization measurements.

**Symmetry analysis**

In order to proof the consistency of the proposed magnetic structures with the symmetry requirements of space group $I\bar{4}2d$ a symmetry analysis of the magnetic structure has been conducted using the BASIREPS tool included in the *FullProf* Suite package. For the Mn1 site this results in two 1-dim irreducible representations (IRs) $\Gamma1$ and $\Gamma2$ contained one time, two 1-dim IRs $\Gamma3$ and $\Gamma4$ contained two times, and one 2-dim IR $\Gamma5$ contained three times. Decomposition of the magnetic representation for the Mn2 site yields two 1-dim IRs $\Gamma1$ and $\Gamma3$ contained one time and one 2-dim IR $\Gamma5$ contained two times. We have tested all the possible combinations of the IRs for Mn1 and Mn2 sites and concluded that only the combination $\Gamma3$ (Mn1) and $\Gamma3$ (Mn2) gives a satisfactory agreement with the experimentally observed magnetic structure above $T_{SR}$. The corresponding model contains three free parameters that represent the $\mu_z$ component on the Mn2 site — $\mu_z(Mn2)$, alternating in-plane components $\mu_x$ and $\mu_y$ on the Mn1 site (in $y$, $-y$, $-x$, $x$ sequence) — $\mu_{xy}(Mn1)$, and the $\mu_z$ component on the Mn1 site — $\mu_z(Mn1)$. We assumed that the parameter $\mu_{xy}(Mn1)$ is zero, which in essence gives the model of a collinear ferromagnetic or ferrimagnetic structure. As follows from our analysis, there exists no single-IR model that can describe the magnetic structure below $T_{SR}$. Therefore, the models that include a superposition of two IRs on each Mn-sublattice were considered. Among those models, the combination of $\Gamma3 + \Gamma5$ (Mn1) and $\Gamma3 + \Gamma5$ (Mn2) was found to give the best agreement with the collected diffraction pattern. The model implies a set of thirteen parameters, from which only five parameters are assumed to have finite values and two parameters are constrained to be oppositely equal. This restricts the model to the set of four effective parameters: $\mu_z(Mn1)$, $\mu_x(Mn1)$, $\mu_z(Mn2)$, and $\mu_x(Mn2)$. The proposed model describes a non-collinear coplanar magnetic structure.

**Neutron studies on Mn$_{1.4}$Pt$_{1-y}$Rh$_y$Sn**

In contrast to the Mn$_{1.4}$Pt$_{0.9}$Pd$_{0.1}$Sn data, the neutron powder diffraction patterns of Mn$_{1.4}$Pt$_{1-y}$Rh$_y$Sn ($y = 0.4$, 0.6, 0.7 and 0.8) were collected with reduced statistics and also within a more restricted $2\theta$ range. This led to failure of the model with four independent parameters. The refinement of the data with the full four-parameter model resulted in some unphysical temperature evolution of the $\mu_x$(Mn2) component in case of $y = 0.4$, and also to unphysical behavior of $\mu_z$(Mn1) and $\mu_z$(Mn2) for higher substitutions $y = 0.6, 0.7, 0.8$. Therefore, we simplified our low-temperature model to three parameters: $\mu_z$(Mn1), $\mu_x$(Mn1), and $\mu_z$(Mn2), and to two parameters: $\mu_x$(Mn1) and $\mu_x$(Mn2), for the magnetic structure of $y = 0.4$ and $y \geq 0.6$ compounds, respectively.

The results of the refinement of $Mn_{1.4}Pt_{0.6}Rh_{0.4}Sn$ can be summarized as follows. At temperatures below $T_{SR}$ the magnetic moments of the Mn2 sublattice are oriented along the [001] direction, whereas the magnetic moment of the Mn1 site lies in the (010) plane. The relation $\mu_x$(Mn1) $> \mu_z$(Mn1) holds, thus the Mn1 moment is more inclined toward [100]. At $T =$ 1.5 K we obtained $\mu_x$(Mn1) $= 3.0(1)$ $\mu_B$, $\mu_z$(Mn1) $= 0.4(2)$ $\mu_B$, and $\mu_z$(Mn2) $= 3.0(1)$ $\mu_B$. The canting angle (the angle between Mn1 and Mn2 moments) calculated from the refined parameters was ~80°, which is close to the canting angle of the compound $Mn_{1.4}Pt_{0.9}Pd_{0.1}Sn$. Therefore, within the simplified model we cannot determine any significant difference in the magnetic structures of the $x = 0.1$ and $y = 0.4$ compounds. Fine details in the change of the non-collinear spin structure of $Mn_{1.4}Pt_{1-y}Rh_ySn$ ($y = 0 – 0.4$) should be addressed in future studies.

In contrast to the $y = 0.4$ compound, we observe a drastic change of the magnetic structure in the compound with 60% Rh substitution ($y = 0.6$). The model of a collinear ferrimagnetic structure with the net moment along [100] gives plausible agreement with the data. The refined parameters at low temperature yield $\mu_x$(Mn1) $= 2.7(1)$ $\mu_B$ and $\mu_x$(Mn2) $= -1.4(1)$ $\mu_B$. The inclusion of the additional magnetic components $\mu_z$(Mn1) and $\mu_z$(Mn2) does not cause any improvement in the residual. It is worth to mention that $z$ components of the magnetic moments as small as 0.5 $\mu_B$ could remain undetected within the limited statistics of the collected data. Hence, one cannot conclude whether $Mn_{1.4}Pt_{0.4}Rh_{0.6}Sn$ is a fully collinear ferrimagnet or there exists some small canting of the magnetic moments away from the [100] crystallographic direction. The model of the collinear ferrimagnet successfully describes the diffraction patterns at all the temperatures. We also observed no change in the magnetic structure between compounds with $y = 0.6, 0.7, 0.8$. We present the neutron diffraction patterns and refinement results for $y = 0.4$ and 0.6 at 1.5 K in Fig. S2.

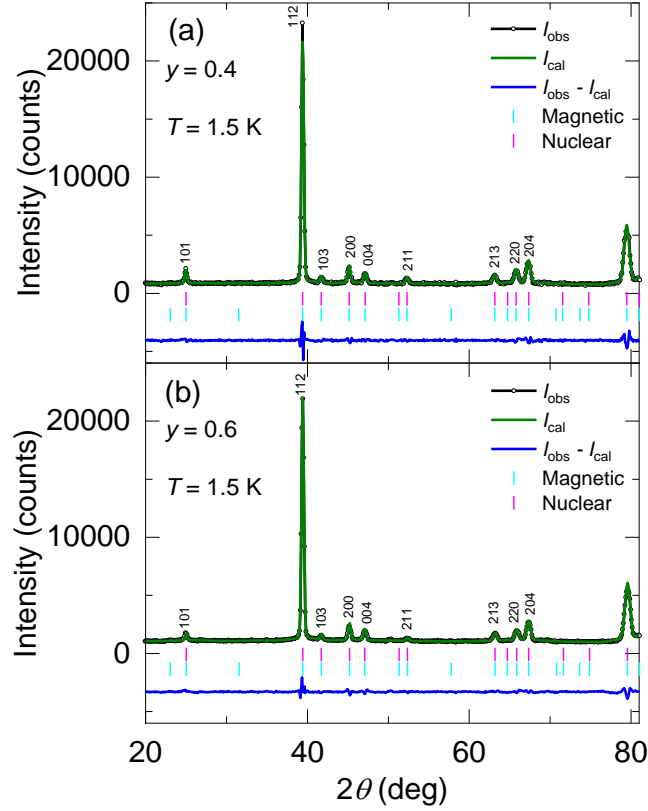

FIG. S2. Rietveld refinement results for $Mn_{1.4}Pt_{1-y}Rh_ySn$ with (a) $y = 0.4$ and (b) $y = 0.6$ at 1.5 K.

### III. Magnetic characterization

Temperature and field dependent magnetization measurements were performed using a vibrating sample magnetometer (MPMS3, Quantum Design). The DC $M(T)$ measurements were carried out from 2 to 400 K in the zero field cooling (ZFC), field cooled cooling (FCC) and field cooled warming (FCW) modes using a field of 0.01 and 0.1 T. The AC susceptibility measurements were taken from 2 to 400 K using a small AC driving field of 5 Oe and a frequency of 18 Hz in the absence of a DC bias field. Magnetic isotherms $M(H)$ were measured in the field range of −7 to +7 T at different temperatures between 2 and 300 K. The field dependent AC susceptibility measurements were carried out using a small AC driving field of 5 Oe and a frequency of 486 Hz in the field range of ±5 T.

We present the magnetic properties of $Mn_{1.4}Pt_{0.9}Pd_{0.1}Sn$ in Fig. S3. The $T_C$ (~392 K) and $T_{SR}$ (~133 K) values obtained from the positions of the $\chi'$-maxima are consistent with the neutron study. The magnetic isotherms have different shapes below (2 and 120 K) and above (150 and 200 K) $T_{SR}$ with a negligible coercivity. The corresponding field dependent $\chi'$ demonstrates a hysteretic behavior at temperatures below $T_{SR}$. At 2 K, this feature is present in the field range of ±0.5 T. It decreases with temperature and disappears above $T_{SR}$. The bifurcation of the Hall resistivity (Fig. 5(a)) and the irreversibility of $\chi'$ at different field sweep directions match well. This further supports the formation of a skyrmion-like spin texture [1,2] which gives rise to the finite value of the Hall resistivity at zero field despite of the zero magnetization.

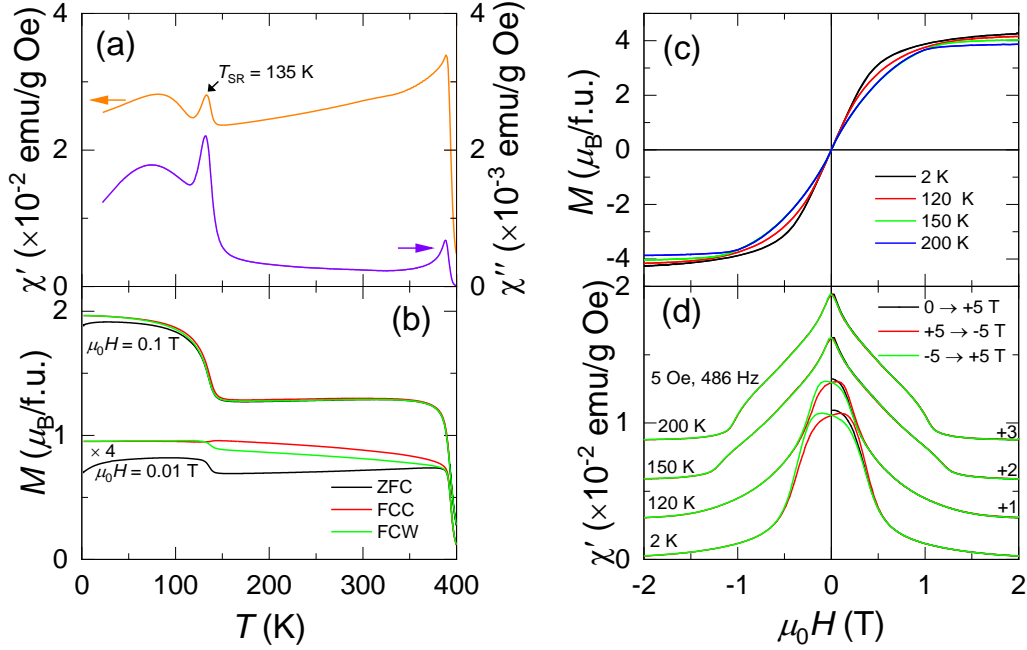

FIG. S3. Magnetization measurements of $Mn_{1.4}Pt_{0.9}Pd_{0.1}Sn$. (a) The real ($\chi'$) and imaginary ($\chi''$) parts of the AC susceptibility as a function of temperature. (b) Temperature dependent magnetization $M(T)$ plots at 0.01 and 0.1 T in ZFC, FCC and FCW modes. (c) Magnetic isotherms and (d) field dependent $\chi'$ at 2, 120, 150 and 200 K. +1 represents the displacement of the $\chi'$ plot by 0.01 emu/g Oe and so on.

We depict the temperature dependent magnetization measurements of the synthesized compound series $Mn_{1.4}Pt_{1-x}Pd_xSn$ and $Mn_{1.4}Pt_{1-y}Rh_ySn$ in Fig. S4. We observe a spin reorientation transition < 150 K in all the samples except for $y > 0.5$. A broad transition for $y = 0.6$ below 200 K, becomes less pronounced with increasing Rh substitution. We present the magnetic isotherms $M(H)$ of Rh substituted samples at 2 K in Fig. S5.

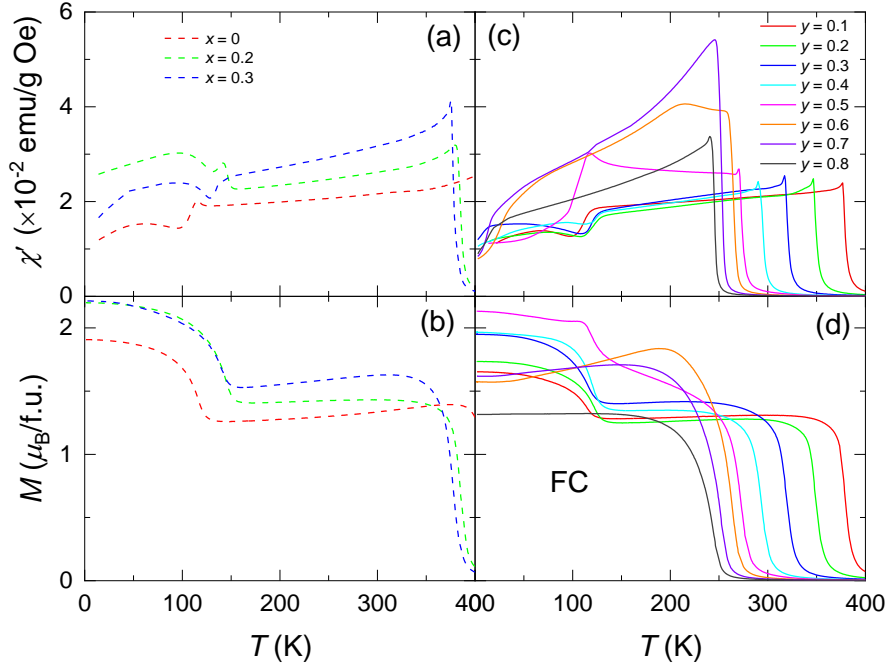

FIG. S4. Temperature dependent real part $\chi'$ of the AC susceptibility and magnetization $M(T)$ for $Mn_{1.4}Pt_{1-x}Pd_xSn$ (a,b) and $Mn_{1.4}Pt_{1-y}Rh_ySn$ (c,d).

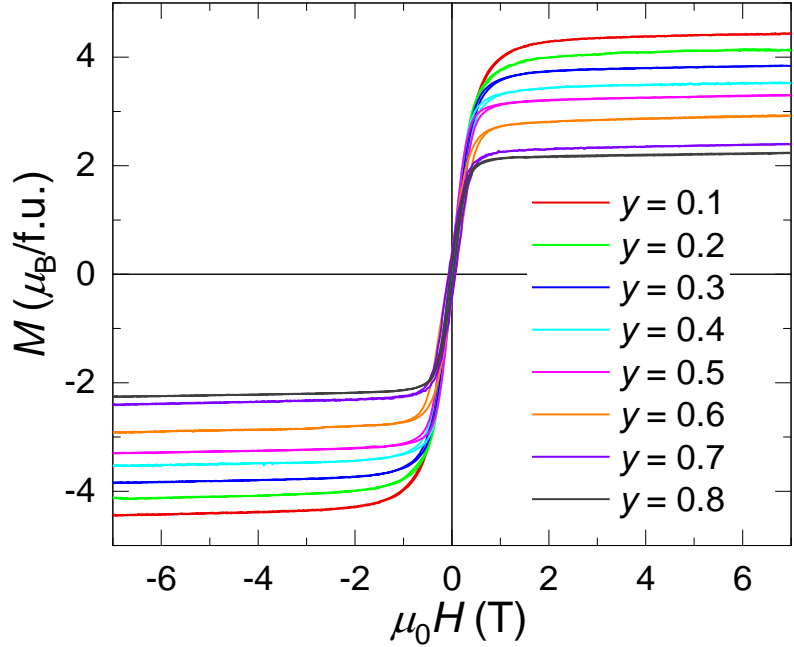

FIG. S5. Magnetic isotherms of $Mn_{1.4}Pt_{1-y}Rh_ySn$.

To quantify the magnetocrystalline anisotropy in the polycrystalline ferromagnetic compounds, we used the law of approach to saturation [3] which can be expressed as $M = M_S(1 - a_1/H - a_2/H^2 - \cdots) + \chi H$, where the symbols $M$, $M_S$, $\chi$ and $H$ denote the magnetization, saturation magnetization, high field susceptibility and internal field, respectively. Here, $a_1$ and $a_2$ are constants. The third term in the equation corresponds to the magnetocrystalline anisotropy that has a dominating part at higher fields. For uniaxial crystals, $a_2 = 4K_1^2/15M_s^2$, where $K_1$ is the magnetocrystalline anisotropy constant [3]. We calculated the magnetocrystalline anisotropy constant $K_1$ at 200 K (in the collinear ferromagnetic phase where spins align along the tetragonal axis) using the slope and intercept of the straight line of $M$ vs $1/H^2$ plots for the fields above 5 T. The calculated magnetocrystalline anisotropy constant $K_1$ and the anisotropy field $H_K = 2K_1/M_S$ at 200 K are shown in Fig. S6. The anisotropy constant of $Mn_{1.4}PtSn$ is 7.2 Merg/cm$^3$ (7.2 × 10$^5$ J/m$^3$), which is of the same order of magnitude as experimentally derived for $Mn_2PtSn$ [4]. The Pd substitution leads a slight decrease in $K_1$ and $H_K$. In contrast, $K_1$ increases to $y = 0.5$ for Rh substitution which is consistent with the decreasing $c/a$ ratio (Fig. 1) of the compounds. The used formula does not apply for $y = 0.6, 0.7$ and $0.8$ as in this composition range a ferrimagnetic structure was found (see section **II**).

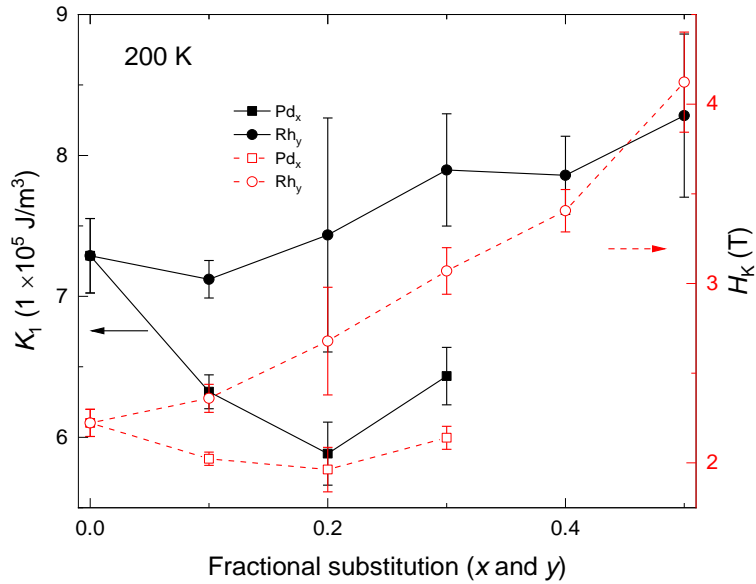

FIG. S6. Calculated magnetocrystalline anisotropy constant $K_1$ and anisotropy field $H_K$ at 200 K of $Mn_{1.4}Pt_{1-x}Pd_xSn$ and $Mn_{1.4}Pt_{1-y}Rh_ySn$.

## IV. Transport measurements

The electrical resistivity $\rho_{xx}$, measured by lowering the temperature from 300 to 2 K is shown in Fig. S7(a). The resistivity decreases with decreasing the temperature in all the samples signifying their metallic character. The parent compound, all Pd substituted and the Rh substituted samples with $y < 0.5$ have $T_C$'s above room temperature. The samples with higher levels of Rh substitution ($y > 0.4$) show an anomaly around their $T_C$ corresponding to the transformation from the paramagnetic disordered phase to the ferro-/ferrimagnetic ordered phase. Another anomaly appears at the temperature corresponding to the spin reorientation transition. It can be clearly seen from the $\rho_{xx}(T)$ plots that the rate of decrement enhances below $T_{SR}$. We observed a general trend of increase in the resistivity as well as in the residual resistivity at 2 K with the chemical substitution (increasing $x$ and $y$).

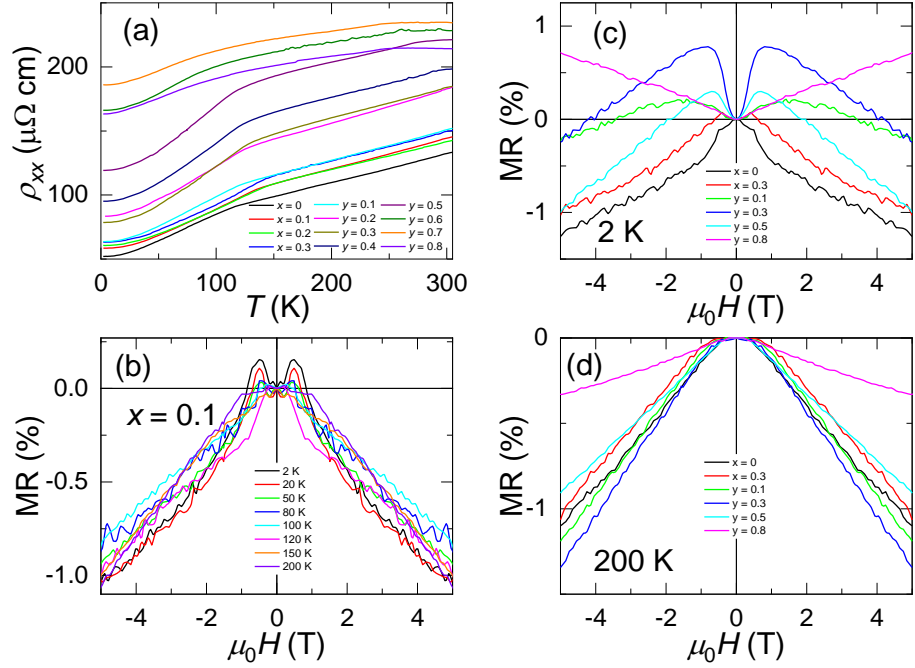

FIG. S7. Transport properties of Mn$_{1.4}$Pt$_{1-x}$Pd$_x$Sn and Mn$_{1.4}$Pt$_{1-y}$Rh$_y$Sn. (a) Resistivity $\rho_{xx}$ as a function of temperature in zero field. (b) Transverse magnetoresistance (MR) in % at various temperatures for $x = 0.1$. The MR for different $x$ and $y$ is depicted in (c) at 2 K and (d) at 200 K, respectively.

We measured the transverse magnetoresistance (MR) in the field range of ±5 T with varying the temperature. The MR was symmetrized in order to remove any contribution of the Hall resistivity by taking the average of corresponding positive and negative field values. The MR given in percentage was calculated using the equation MR(%) = $\frac{\rho(\mu_0 H) - \rho_0}{\rho_0} \times 100$ where $\rho(\mu_0 H)$ and $\rho_0$ are the transverse resistivity and zero field resistivity, respectively. We show the MR (%) for $x = 0.1$ at different temperatures in Fig. S7(b). The MR at 2 K is positive for fields up to 0.5 T and becomes negative for higher fields. This anomaly in MR decreases with rise in temperature and vanishes above $T_{SR}$ which seems to be correlated with the Hall and the AC susceptibility measurements. Fig. S7(c) and S7(d) depict the MR (%) for $x = 0$ and 0.3, and $y = 0.1$, 0.3, 0.5 and 0.8 at 2 K ($T < T_{SR}$) and 200 K ($T > T_{SR}$), respectively.

We present the method of extraction of the topological Hall resistivity $\rho^T$ in Fig. S8 (as described in Ref. [5]) taking as an example the Hall resistivity, MR and magnetization data for $x = 0.3$ at 50 K. The Hall resistivity can be expressed as the sum of three contributing components by the relation $\rho_{yx} = \rho^N + \rho^A + \rho^T$, where $\rho^N$, $\rho^A$ and $\rho^T$ denote the normal, anomalous and topological Hall resistivity, respectively. The normal part is proportional to the applied magnetic field and defined as $\rho^N = R_0 \mu_0 H$. Here, $R_0$ is the normal Hall coefficient. On the other hand, the anomalous Hall resistivity is proportional to the magnetization $M$ and can be described by the relation $\rho^A = b\rho_{xx}^2 M$, where $b$ is the proportionality constant. At higher fields where the magnetization nearly saturates, $\rho^T = 0$. Therefore, $\rho_{yx} = R_0 \mu_0 H + b\rho_{xx}^2 M$ and the plot of $\rho_{yx}/\mu_0 H$ ($y$-axis) vs $\rho_{xx}^2 M/\mu_0 H$ ($x$-axis) gives a straight line. A linear fit gives the slope $b$ and intercept $R_0$ values from which the topological Hall resistivity can be derived by the formula $\rho^T = \rho_{yx} - R_0 \mu_0 H - b\rho_{xx}^2 M$.

In general, the anomalous Hall resistivity is a function of both $\rho_{xx}$ and $\rho_{xx}^2$ and has the form $\rho^A = (a\rho_{xx} + b\rho_{xx}^2)M$, where $a$ and $b$ are anomalous Hall coefficients. MR measurements show (see Fig. S7) that the variation of $\rho_{xx}$ is very small (within 1.5 %). Therefore, we can take $\rho^A = R_S M$, where $R_S = a\rho_{xx} + b\rho_{xx}^2$ is considered as a constant. Now we extract topological Hall resistivity assuming that $\rho_{xx}$ is a constant (described in Fig. S9). The constants $R_0$ and $R_S$ are derived by the straight line fitting of the equation $\rho_{yx} = R_0 \mu_0 H + R_S M$. The extraction of topological Hall resistivity does not show any significant difference on the form of anomalous Hall resistivity due to the small variation of $\rho_{xx}$.

In Fig. S10 we present the log-log plot of $\rho_{xx}$ ($x$-axis) vs $\rho^A$ ($y$-axis) at various temperatures. A straight-line feature with slope greater than 1.6 is present in the parent ($x = 0$) and all Pd substituted compounds at $T \leq 50$ K. This indicates the dominating character of the $\rho_{xx}^2$ term in the anomalous Hall resistivity at lower temperatures. However, the slope decreases at higher temperatures. We found a drastic change in the straight-line feature nearby the spin reorientation transition. On the other hand, the slope is less than 1.5 for moderately substituted Rh compounds $y \leq 0.3$ at $T \leq 50$ K. There is no clear indication of scaling behavior of $\rho^A$ on $\rho_{xx}$ for $y \geq 0.4$. Therefore, for simplification we used the quadratic term of $\rho_{xx}$ in the anomalous Hall resistivity $\rho^A$ to extract the topological Hall effect for all compounds.

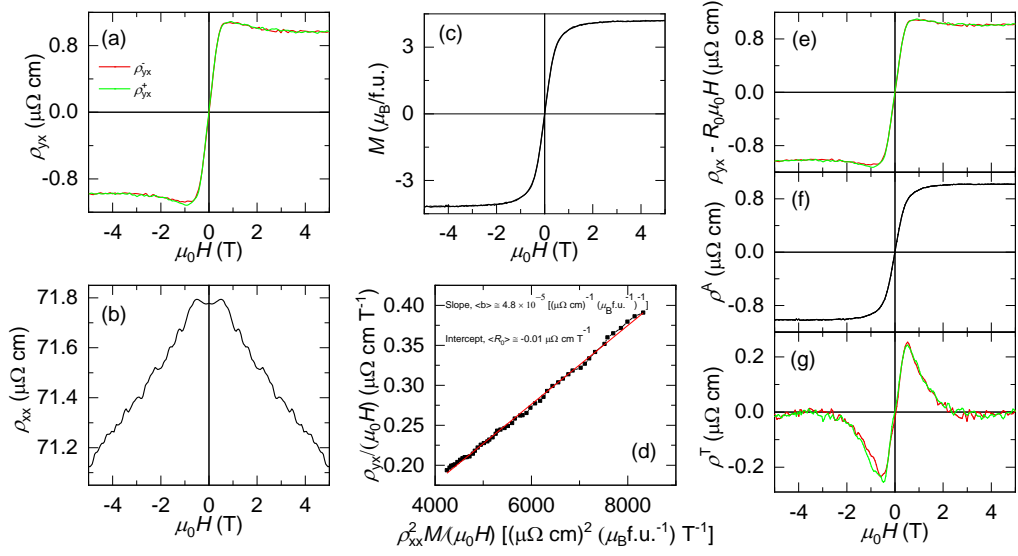

FIG. S8. Method of extraction of topological Hall resistivity. (a) Hall resistivity $\rho_{yx}$, (b) MR and (c) magnetic isotherm $M(H)$ for $x = 0.3$ at 50 K, (d) $\rho_{yx}/\mu_0 H$ vs $\rho_{xx}^2 M/\mu_0 H$ gives $R_0$ and $b$, (e) $\rho_{yx} - R_0\mu_0 H$, (f) $\rho^A (= b\rho_{xx}^2 M)$ and (g) $\rho^T (= \rho_{yx} - R_0\mu_0 H - b\rho_{xx}^2 M)$.

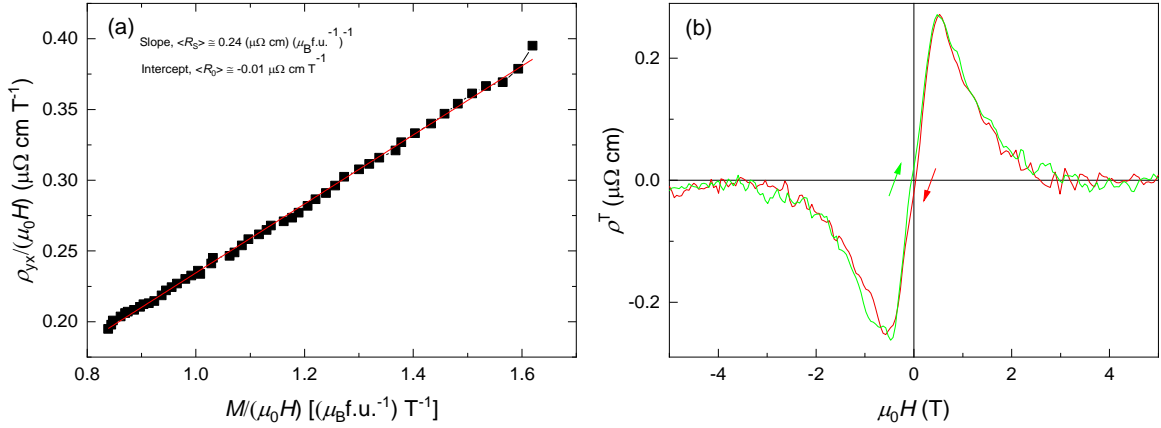

FIG. S9. Extraction of topological Hall resistivity for $x = 0.3$ at $T = 50$ K assuming that $\rho_{xx}$ is a constant and $\rho^A = R_S M$. (a) $\rho_{yx}/\mu_0 H$ vs $M/\mu_0 H$ gives $R_0$ and $R_S$. (b) $\rho^T (= \rho_{yx} - R_0\mu_0 H - R_S M)$.

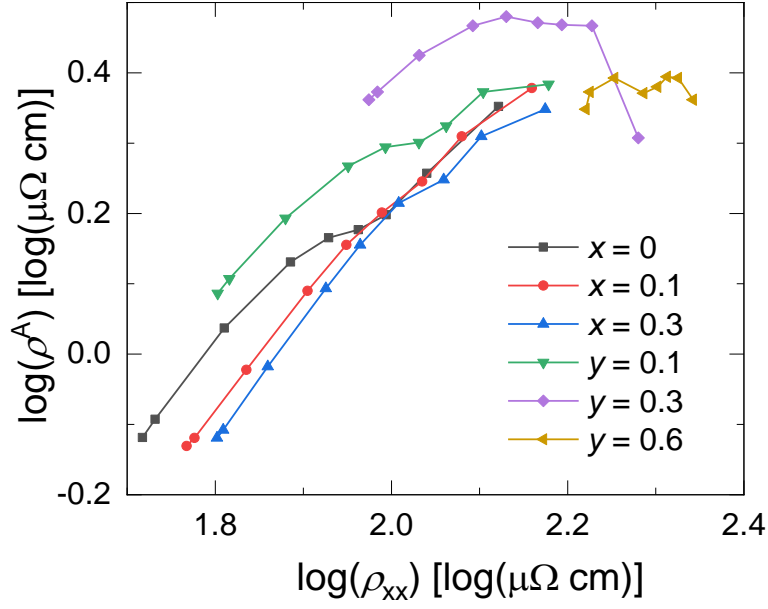

FIG. S10. Log-log plot of $\rho_{xx}$ (x-axis) vs $\rho^A$ (y-axis) at various temperatures.

In Fig. S11 we show the decrement of topological Hall resistivity, $\rho^T$ above the spin reorientation transition by considering an example of $x = 0.1$ ($T_{SR} = 135$ K). At 150 K ($T > T_{SR}$), we still observe a considerable difference between $\rho_{yx} - R_0\mu_0 H$ and magnetization scaled anomalous Hall resistivity, $\rho^A$ which results in a finite value of $\rho^T$. However, the reversal of Hall resistivity is only present at temperatures $T < T_{SR}$. On the other hand, at 300 K ($T \gg T_{SR}$) $\rho_{yx} - R_0\mu_0 H$ and $\rho^A$ almost superimposes and there is no indication of topological Hall effect.

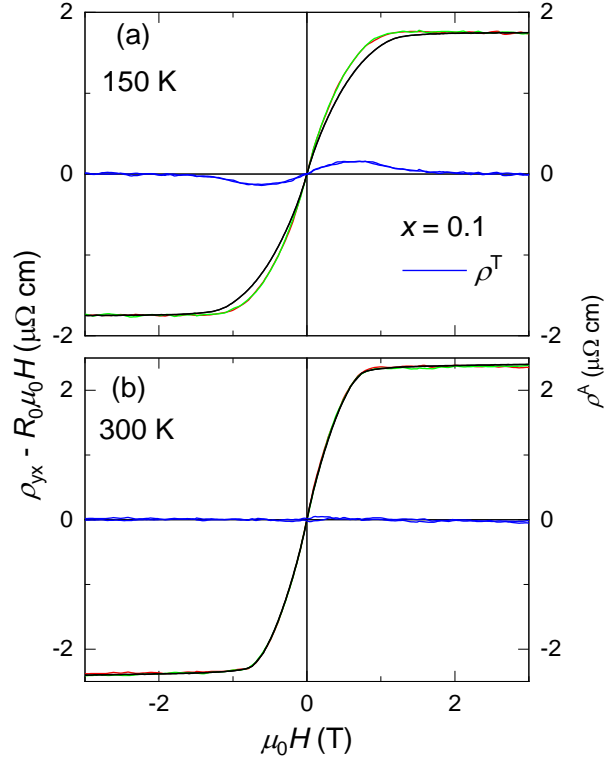

FIG. S11. (a) Corrected Hall resistivity $\rho_{yx} - R_0\mu_0 H$ (left axis, red and green lines) and the anomalous Hall resistivity $\rho^A (\propto M)$ (right axis, black lines) at (a) 150 K and (b) 300 K of $Mn_{1.4}Pt_{0.9}Pd_{0.1}Sn$ ($x = 0.1$). The topological Hall resistivity, $\rho^T (= \rho_{yx} - R_0\mu_0 H - R_S M)$ is shown by the blue colour.

The Hall and topological Hall resistivity of $x = 0$, 0.2 and 0.3 is shown in Fig. S12 displaying similar features as that of $x = 0.1$. We show the Hall resistivity of $Mn_{1.4}Pt_{1-y}Rh_y Sn$ in Fig. S13. Fig. S14 demonstrates that a finite but comparatively smaller topological Hall effect (THE) is also observed for $y = 0.7$. The present effect may be attributed to a small canting angle between moments of Mn1 and Mn2 atoms in an essentially ferrimagnetic configuration or it may be an artifact due to the error generated in the extraction of the topological Hall resistivity. However, there is no indication of THE for $y = 0.8$. The anomalous Hall conductivity $\sigma^A$ (shown in Fig. S15) as a function of temperature demonstrates a significant enhancement below $T_{SR}$. The compound with $y = 0.6$ has a slight increase in conductivity, however no enhancement is observed for $y = 0.7$ and 0.8. This is consistent with the scaling of $\sigma^A$ by the magnetization $M$.

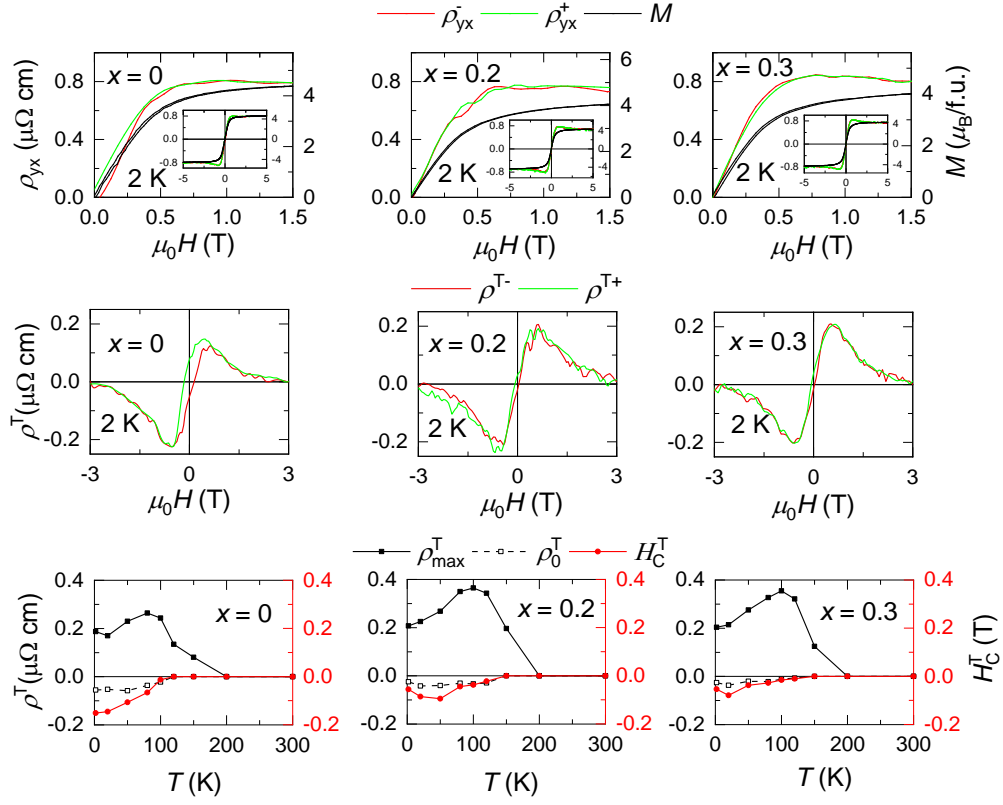

FIG. S12. Hall resistivity data of Mn$_{1.4}$Pt$_{1-x}$Pd$_x$Sn. Top panel: Hall resistivity $\rho_{yx}$ (left axis) and magnetization (right axis) for $x = 0$, 0.2 and 0.3 between 0 and 1.5 T at 2 K. The inset shows full hysteresis loops between −5 and +5 T. Middle panel: The corresponding topological Hall resistivity $\rho^T$ at 2 K. Bottom panel: Temperature dependent variation of the maximum and zero field value and of the coercivity of the topological Hall resistivity.

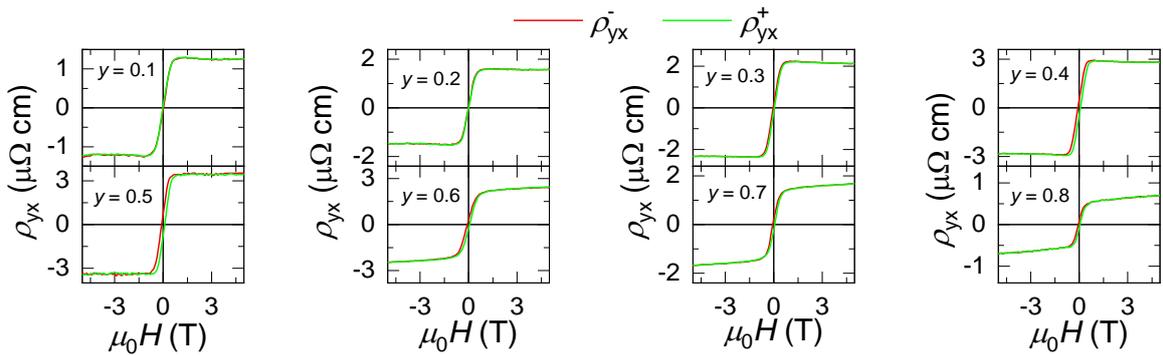

FIG. S13. The experimental Hall resistivity of Mn$_{1.4}$Pt$_{1-y}$Rh$_y$Sn.

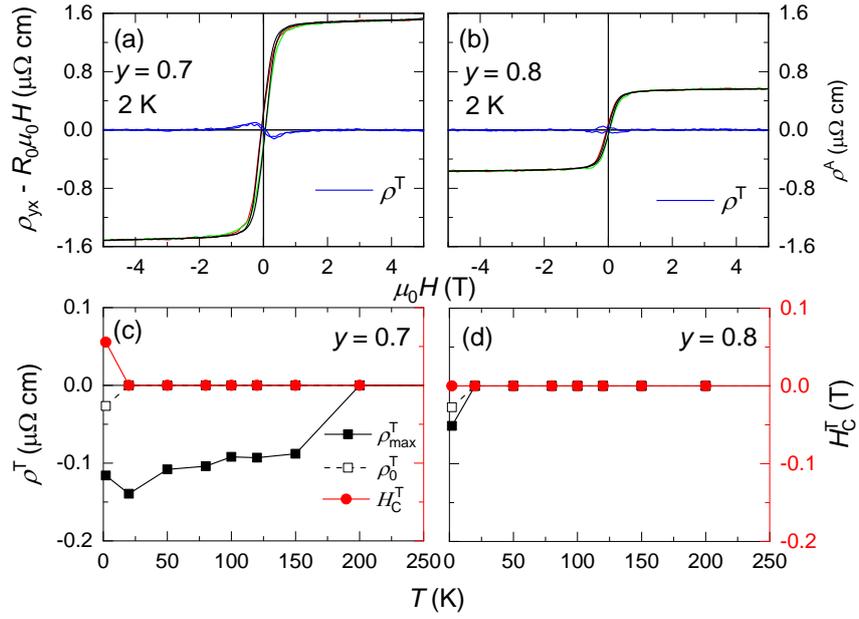

FIG. S14. Corrected Hall resistivity $\rho_{yx} - R_0\mu_0 H$ (left axis, red and green lines) and the anomalous Hall resistivity $\rho^A (\propto M)$ (right axis, black lines) at 2 K for (a) $y = 0.7$ and (b) $y = 0.8$. The topological Hall resistivity, $\rho^T (= \rho_{yx} - R_0\mu_0 H - R_S M)$ is shown by the blue colour. Temperature dependent variation of the maximum and zero field value and of the coercivity of the topological Hall resistivity for (c) $y = 0.7$ and (d) $y = 0.8$.

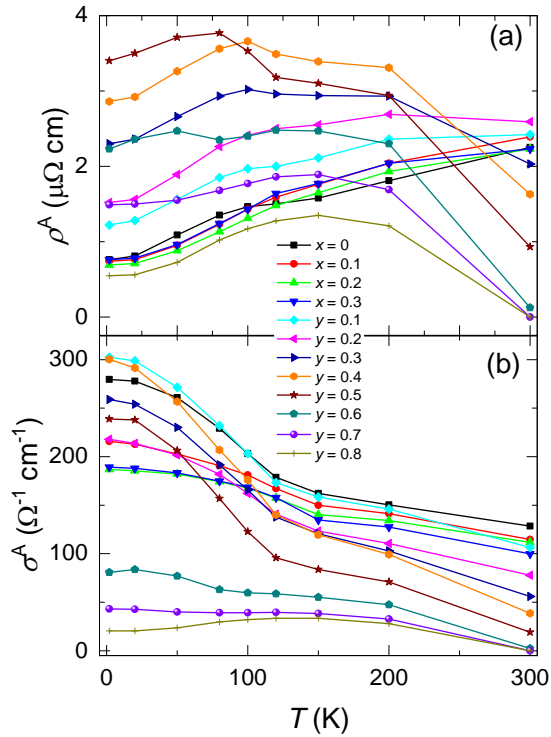

FIG. S15. Anomalous Hall resistivity and conductivity of $Mn_{1.4}Pt_{1-x}Pd_xSn$ and $Mn_{1.4}Pt_{1-y}Rh_ySn$.

* claudia.felser@cpfs.mpg.de

References:

[1]  Y. Tokunaga, X. Z. Yu, J. S. White, H. M. Rønnow, D. Morikawa, Y. Taguchi, and Y. Tokura, Nat. Commun. **6**, 7638 (2015).

[2]  K. Karube, J. S. White, N. Reynolds, J. L. Gavilano, H. Oike, A. Kikkawa, F. Kagawa, Y. Tokunaga, H. M. Rønnow, Y. Tokura, and Y. Taguchi, Nat. Mater. **15**, 1237 (2016).

[3]  S. V. Andreev, M. I. Bartashevich, V. I. Pushkarskya, V. N. Maltsev, L. A. Pamyatnykh, E. N. Tarasov, N. V. Kudrevatykh, and T. Goto, J. Alloys Compd. **260**, 196 (1997).

[4]  Y. Huh, P. Kharel, A. Nelson, V. R. Shah, J. Pereiro, P. Manchanda, A. Kashyap, R. Skomski, and D. J. Sellmyer, J. Phys. Condens. Matter **27**, 076002 (2015).

[5]  see Ref. 7 in the Main Text.